\documentclass[a4paper,12pt, epsfig]{article}
\def\letter{0}\def\pr{0}
\pdfoutput=1 
\usepackage{epsfig}
\usepackage{epstopdf}
\usepackage{graphicx}
\usepackage{ifthen}
\usepackage{ulem}

\usepackage{amsmath}
\usepackage[psamsfonts]{amssymb}
\usepackage{euscript}

\usepackage{soul}

\usepackage{latexsym}
\usepackage[arrow,matrix,curve]{xy}

\usepackage[hypertexnames=false]{hyperref}
\hypersetup{
    colorlinks=true,
    linkcolor=blue,
    filecolor=magenta,   
    citecolor=black,   
    urlcolor=cyan,
    }

\newskip\humongous \humongous=0pt plus 1000pt minus 1000pt

\newif\ifdtup

\def\,{\hspace{-.1cm}}
\def\hsp{,\hspace{.7cm}}

\def\fc#1#2 {\frac{n}{q}#1\frac{n}{q}#2}

\renewcommand{\cos}{\textrm{cos}}
\renewcommand{\sin}{\textrm{sin}}

\renewcommand{\sinh}{\textrm{sinh}}
\renewcommand{\cosh}{\textrm{cosh}}
\renewcommand{\tanh}{\textrm{tanh}}
\newcommand{\sech}{\textrm{sech}}
\newcommand{\csch}{\textrm{csch}}

\renewcommand{\theequation}{\arabic{section}.\arabic{equation}}
\renewcommand{\(}{\begin{equation}}
\renewcommand{\)}{end{equation} \vspace{-.05in}\linebreak}

\newcounter{saveeqn}
\newcounter{savealpheqn}

\newcommand{\alpheqn}{\setcounter{saveeqn}{\value{equation}}%
  \stepcounter{saveeqn}\setcounter{equation}{0}%
  \renewcommand{\theequation}{\mbox{\arabic{section}.\arabic{saveeqn}
\alph{equation}}}
  \renewcommand{\)}{\end{equation}}}
\def\part#1{\frac{\partial}{\partial{#1}}}%
\def\group#1{\refstepcounter{equation}\setcounter{saveeqn}
 {\value{equation}}%
  \label{#1}\setcounter{equation}{0}%
\renewcommand{\theequation}{\mbox{\arabic{section}.\arabic{saveeqn}
\alph{equation}}}
  \renewcommand{\)}{\end{equation}}}
\newcommand{\reseteqn}{\setcounter{equation}{\value{saveeqn}}%
  \renewcommand{\theequation}{\arabic{section}.\arabic{equation}}%
  \renewcommand{\)}{\end{equation}}}

\newcommand{\aalpheqn}{\setcounter{saveeqn}{\value{equation}}%
  \stepcounter{saveeqn}\setcounter{equation}{0}%
  \renewcommand{\theequation}{\mbox{
        \Alph{subsection}.\arabic{saveeqn}\alph{equation}}}
   \renewcommand{\)}{\end{equation}}}
\newcommand{\areseteqn}{\setcounter{equation}{\value{saveeqn}}%
  \renewcommand{\theequation}{\Alph{subsection}.\arabic{equation}}%
  \renewcommand{\)}{\end{equation}}}

\renewcommand{\thefootnote}{\alph{footnote}}
\renewcommand{\(}{\begin{equation}}
\renewcommand{\)}{\end{equation}}
\newcommand{\ba}{\begin{eqnarray}}
\newcommand{\ea}{\end{eqnarray}}
\newcommand{\cbp}{\mathop{\vtop{\ialign{##\crcr
   $\hfil\displaystyle{}\hfil$\crcr\noalign{\kern-13pt\nointerlineskip}
   \BIG{)}\hskip0pt\crcr\noalign{\kern3pt}}}}}
\newcommand{\pa}{\mathop{\vtop{\ialign{##\crcr

$\hfil\displaystyle{\oplus}\hfil$\crcr\noalign{\kern+1pt\nointerlineskip
}
   \hspace{.08in}$^{\alpha=0}$\hskip6pt\crcr\noalign{\kern3pt}}}}}
\renewcommand{\hsp}{,\hspace{.3in}}
\newcommand{\p}{^\prime}
\newcommand{\pp}{^{\prime\prime}}

\catcode`\@=11
\def\vereq#1#2{\lower3pt\vbox{\baselineskip1.5pt \lineskip1.5pt
\ialign{$\m@th#1\hfill##\hfil$\crcr#2\crcr\sim\crcr}}}
\catcode`\@=12

\renewcommand{\(}{\begin{equation}}
\renewcommand{\)}{\end{equation}}

\def\hk{{\hat k}}
\def\pin#1{\int \frac{d#1}{2\pi}}
\def\ppin#1{\int\hspace{-17pt}\sum \frac{d#1}{2\pi}}

\def\dint{\int\hspace{-12pt}\sum }

\newcommand{\beas}{\begin{eqnarray*}}
\newcommand{\eeas}{\end{eqnarray*}}

\newcommand{\bquo}{\begin{quote}}
\newcommand{\enqu}{\end{quote}}

\def\lim#1{\stackrel{\rm{lim}}{{}_{#1}}}
\def\oe#1{\Big\vert_{O(\epsilon^#1)}}

\newcommand{\g}{\mathfrak g}

\def\ch{{\mathcal{H}}}

\def\ok#1{\omega_{k_{#1}}}

\newcommand{\beq}{\begin{equation}}
\newcommand{\eeq}{\end{equation}}
\newcommand{\bea}{\begin{eqnarray}}
\newcommand{\eea}{\end{eqnarray}}

\newskip\humongous \humongous=0pt plus 1000pt minus 1000pt

\newif\ifdtup

\catcode`\@=11

\ifthenelse{\equal{\letter}{0}}{ 

\@addtoreset{equation}{section}
\def\theequation{\arabic{section}.\arabic{equation}}

\def\@normalsize{\@setsize\normalsize{15pt}\xiipt\@xiipt
\abovedisplayskip 14pt plus3pt minus3pt%
\belowdisplayskip \abovedisplayskip
\abovedisplayshortskip \z@ plus3pt%
\belowdisplayshortskip 7pt plus3.5pt minus0pt}

\def\small{\@setsize\small{13.6pt}\xipt\@xipt
\abovedisplayskip 13pt plus3pt minus3pt%
\belowdisplayskip \abovedisplayskip
\abovedisplayshortskip \z@ plus3pt%
\belowdisplayshortskip 7pt plus3.5pt minus0pt
\def\@listi{\parsep 4.5pt plus 2pt minus 1pt
      \itemsep \parsep
      \topsep 9pt plus 3pt minus 3pt}}

\relax

\def\section{\@startsection{section}{1}{\z@}{3.5ex plus 1ex minus  .2ex}{2.3ex plus .2ex}{\large\bf}}

\def\thesection{\arabic{section}}
\def\thesubsection{\arabic{section}.\arabic{subsection}}

\def\appendix{\setcounter{section}{0}
 \def\thesection{Appendix \Alph{section}}
 \def\thesubsection{\Alph{section}.\arabic{subsection}}
 \def\theequation{\Alph{section}.\arabic{equation}}}
\renewcommand{\theequation}{\arabic{section}.\arabic{equation}}

}{
\renewcommand{\theequation}{\arabic{equation}}

} 

\begin{document}
\def\thefootnote{\fnsymbol{footnote}}
\def\thetitle{Normal Modes of the Small-Amplitude Oscillon}
\def\autone{Jarah Evslin}
\def\autthree{Tomasz Roma\'nczukiewicz}
\def\auttwo{Katarzyna S\l awi\'{n}ska}
\def\autfour{Andrzej Wereszczy\'nski}

\def\affc{Institute  of  Theoretical Physics,  Jagiellonian  University,  Lojasiewicza  11,  Krak\'ow,  Poland}
\def\affd{Departamento de Matematica
Aplicada, Universidad de Salamanca, Spain}
\def\affe{International Institute for Sustainability with Knotted Chiral Meta Matter (WPI-SKCM2), Hiroshima University, Higashi-Hiroshima, Hiroshima 739-8526, Japan}
\def\affb{University of the Chinese Academy of Sciences, YuQuanLu 19A, Beijing 100049, China}
\def\affa{Institute of Modern Physics, NanChangLu 509, Lanzhou 730000, China}


\ifthenelse{\equal{\pr}{1}}{
\title{\thetitle}
\author{\autone}
\author{\autthree}
\author{\auttwo}
\affiliation {\affa}
\affiliation {\affb}

}{

\begin{center}
{\large {\bf \thetitle}}

\bigskip

\bigskip


{\large \noindent  \autone{${}^{1,2}$} \footnote{jarah@impcas.ac.cn}, 
\autthree{${}^{3}$} \footnote{tomasz.romanczukiewicz@uj.edu.pl},
 \auttwo{${}^{3}$} \footnote{katarzyna.slawinska@uj.edu.pl},
and \autfour{${}^{3,4,5}$} \footnote{andrzej.wereszczynski@uj.edu.pl}
}


\vskip.7cm

1) \affa\\
2) \affb\\
3) \affc\\
4) \affd\\
5) \affe\\

\end{center}

}

\begin{abstract}
\noindent
Consider a classical (1+1)-dimensional oscillon of small amplitude $\epsilon$.  To all orders in $\epsilon$, the oscillon solution is exactly periodic.  We study small perturbations of such periodic configurations.  These perturbations are themselves periodic up to a monodromy matrix.  We explicitly find the eigenvectors of the monodromy matrix, which are the analogues of normal modes for oscillons.   Dashen, Hasslacher and Neveu used such eigenvectors to quantize the sine-Gordon breather, and we suspect that they will be necessary to quantize the oscillon.  Our results, regardless of the chosen model, suggest that low amplitude oscillons do not reflect small amplitude radiation.

\end{abstract}

%
\setcounter{footnote}{0}
\renewcommand{\thefootnote}{\arabic{footnote}}

\ifthenelse{\equal{\pr}{1}}
{
\maketitle
}{}

\section{Introduction}

Oscillons \cite{osc76,gleiser94, G-cosm, GS} are long-lived, quasi-periodic, soliton-like excitations in non-linear field theories. While genuine (large amplitude) oscillons are rather very complicated objects with characteristic double frequency bahaviour and related amplitude modulations, oscillons with a small amplitude $\epsilon$ seem to be much simpler. Indeed, they are, to all orders in $\epsilon$, periodic \cite{ks74,fodor}.  This implies \cite{pars} that small perturbations about oscillons are also periodic, up to the action of a monodromy matrix.

In Ref.~\cite{dhnsg} the periodic Sine-Gordon soliton was quantized, using the properties of the monodromy matrix and its eigenvectors.  The eigenvectors played the role that was played by normal modes in their earlier quantization of the kink  \cite{dhn2}.  In the case of quantum kinks, the solutions of these normal modes reproduce the reflection and transmission coefficients for kink-radiation scattering at low amplitudes in classical field theory.  Our goal is to do the same for the oscillon, we are interested in both oscillon-radiation scattering and also in the oscillon's quantization.  Thus in the present paper we study the oscillon's monodromy matrix and its eigenvectors.

Perturbations of oscillons have been studied in the past.  Recently, Ref.~\cite{wang22} found a rich structure of perturbations in 2+1 dimensions, as has \cite{van23} in 3+1 dimensions.  However, with an eye towards quantization, we will restrict our attention to the simplest case of (1+1)-dimensional oscillons.  Here, an oscillon whose physical size is $1/\epsilon$ enjoys perturbations at several distinct scales.  At short wavelengths, of order $1/m$ where $m$ is the meson mass, the perturbations are essentially plane waves that scatter with the oscillon.  These were studied in Ref.~\cite{amin10} in a rough approximation where the spatial derivatives of the oscillon, and the resonances that they may lead to, were ignored.  The stability of wavelength $O(1/\epsilon)$ oscillons, whose length scale is similar to the oscillon itself, were investigated in Ref.~\cite{tesi}.  

Perturbations of large amplitude oscillons are a far more complicated issue. However, it has been recently proposed that such perturbed oscillons, which generically exhibit modulations of amplitude, are in fact bound states of unmodulated oscillons \cite{BRSW}. From this perspective, the perturbed oscillon is a nonlinear superposition of several unperturbed (non-modulated) fundamental oscillons just as the modulated two-breather solution of the sine-Gordon theory is a bound state of two single breathers. 

In the present paper, we consider small amplitude oscillons and impose the classical equations of motion up to order $O(\epsilon^3)$, where resonances may first appear and their cancellation places meaningful restrictions on the leading order solutions \cite{fodor}.  We study oscillon perturbations in three regimes.  First we study perturbations of wavenumber $k$ of order $O(m)$ or greater.  These we find are essentially plane waves in a potential of P\"oschl-Teller form, created by the oscillon.  When $k\gtrsim m$, if and only if the potential at the vacuum has a cubic term, at leading order there will be relativistic corrections which imply that the P\"oschl-Teller potential depends explicitly on $k$.  In all cases, we explicitly solve for the leading order perturbations in terms of hypergeometric functions.  We find that the symmetric and antisymmetric solutions are $\pi/2$ out of phase, and so following standard arguments \cite{flugge}, we conclude that in this regime oscillon-radiation scattering is reflectionless for small amplitude radiation.

We also consider a long wavelength regime, where the perturbation has a length scale of order $O(1/\epsilon)$ or more.  Recall that at leading order the oscillon itself has a $\cos(\Omega t+\theta)$ behavior.  This implies that there is a $e^{-i\Omega t}$ positive frequency component and a $e^{i\Omega t}$ negative frequency component.  We find that the perturbations also have two components, one with positive and one with negative frequency.  These are described by two P\"oschl-Teller equations which are coupled inside of the support of the oscillon.  We do not have explicit solutions for these in general.  However, for each positive energy we numerically find an even and an odd unbound state.  These are asymptotically plane-waves, and by numerically calculating their relative phases we find again that the scattering is reflectionless.

In this long-wavelength regime there are also two zero-modes, which satisfy our coupled equations.  One corresponds to the freedom to translate the oscillon, while another corresponds to the phase $\theta$.   On all of the modes mentioned so far, the monodromy matrix can be diagonalized and its eigenvalues are phases, or unity for the zero-modes.  However there is also a perturbation corresponding to a moving oscillon.  Of course, evolving by one period, the kink motion shifts the kink position zero-mode.  Thus the monodromy matrix is not diagonalizable, it is upper triangular.

We also consider intermediate wavenumber perturbations, of order $\sqrt{\epsilon m}$.  For these, the positive and negative frequency perturbations decouple at leading order.  The relativistic corrections also vanish, and so the leading order perturbations are all hypergeometric functions arising from the same P\"oschl-Teller potential.  These are just the nonrelativistic limit of those found for the wavenumber $O(m)$ perturbations, and so again they result in reflectionless scattering.

An important consequence of our analysis is the fact that small amplitude oscillon do not have discrete massive normal modes. 

We begin in Sec.~\ref{fodsez} by reviewing the $\epsilon$-expansion for the oscillon solution \cite{fodor}.  Next in Sec.~\ref{altsez} we consider relativistic perturbations, with wave number $k$ of order $O(m)$ or greater.  In Sec.~\ref{intsez} we consider intermediate scale perturbations and in Sec.~\ref{bassez} we turn to perturbations of the same length scale $O(1/\epsilon)$ as the oscillon itself.  Finally in Sec.~\ref{movsez} we describe the perturbation corresponding to a moving oscillon and the monodromy matrix.

\section{A Low-Amplitude Oscillon} \label{fodsez}

\subsection{The Theory}

Let us consider the classical field theory described by a Hamiltonian
\beq
H=\int dx \ch(x)\hsp
\ch(x)=\frac{\pi(x)^2+\partial_x\phi(x)\partial_x\phi(x)}{2}+\frac{V(g\phi(x))}{g^2}
\eeq
with the zero of $\phi$ and $V$ fixed so that
\beq
V\p(0)=V(0)=0
\eeq
and define
\beq
m^2=V\pp(0) >0.
\eeq
Hamilton's equations are
\beq
\dot \phi(x,t)=\frac{\delta H}{\delta \pi(x)}=\pi(x,t)\hsp
\dot \pi(x,t)=-\frac{\delta H}{\delta \phi(x)}=\partial_x^2\phi(x,t)-\frac{V\p(g\phi(x,t))}{g}
\eeq
so that
\beq
\ddot\phi(x,t)=\partial_x^2\phi(x,t)-\frac{V\p(g\phi(x,t))}{g}
=\partial_x^2\phi(x,t)-m^2\phi(x,t)-\sum_{n=2}^{\infty} \frac{g^{n-1}V^{(n+1)}(0)}{n!}\phi^{n}(x,t)
. \label{eom}
\eeq
We remark that a non-zero mass threshold is not a necessary condition for the existence of oscillons, (it is rather inflection point in the potential or better said an effective potential, see \cite{gleiser2024},\cite{copeland1995} and \cite{For}). However, oscillons in massless models need to have a large amplitude and therefore go beyond the scope of the present work, see \cite{DRSW} for details.  

\subsection{The Oscillon Solution} \label{fodor}
Consider the Ansatz
\beq
\phi(x,t)=f(\epsilon x,t)=\sum_{n=1}^\infty \epsilon^n f_n(\epsilon x,t) \label{expf}
\eeq
where each $f_n$ is of order $O(\epsilon^0)$.  If there is such a solution, this Ansatz contains a broad oscillon $f(\epsilon x,t)$.  The equation of motion can be rewritten as
\beq
\ddot f(\epsilon x,t)=\epsilon^2\partial_{\epsilon x}^2 f(\epsilon x,t)-m^2 f(\epsilon x,t)-\sum_{n=2}^{\infty} \frac{g^{n-1}V^{(n+1)}(0)}{n!}f^n (\epsilon x,t)
. \label{eomf}
\eeq
This slight rewriting has an important consequence.  The only spatial derivative terms are suppressed by order $O(\epsilon^2)$.  This means, that in an $\epsilon$ expansion, one can solve for each $f_n$ using an ordinary differential equation in time at $O(\epsilon^n)$, as its spatial derivative only appears at $O(\epsilon^{n+2})$.  Even at $O(\epsilon^{n+2})$ no partial differential equations appear, merely an ordinary differential equation in time for $f_{n+2}$. Now we insert the expansion (\ref{expf}) into the equation of motion (\ref{eomf}).  

\subsubsection{Order $O(\epsilon)$}

The leading contribution to (\ref{eom}) is at $O(\epsilon)$
\beq
\ddot f_1(\epsilon x,t)\Big\vert_{O(\epsilon^0)}=-m^2 f_1(\epsilon x,t) \label{eqe1}
\eeq
which, for real $f_1$, is solved by
\beq
f_1(\epsilon x,t)=F(\epsilon x)e^{i\Omega t}+F^*(\epsilon x)e^{-i \Omega t}\hsp
\Omega=m+\sum_{n=1}^\infty \epsilon^n \Omega_n\label{om}
\eeq
for some function $F$ of order $O(\epsilon^0)$.   

The $\epsilon$ expansion here is a bit subtle, as in principle $t$ can take any value, and may be of order $O(\epsilon^M)$ for any $M$, as can the time derivatives.  In particular, if $t$ is of order $O(1/\epsilon^2)$ then $O(\epsilon^2)$ corrections to $\Omega$ are of order unity and cannot be relegated to a higher order in the $\epsilon$ expansion.   As a result, our prescription will be as follows.  The time-dependence will be exact, in the sense that the frequency of $f_1$ is exactly $\Omega$ with no $O(\epsilon)$ corrections.  However, $\Omega$ itself will be determined one order at a time.  This approach ensures that solutions to the $\epsilon$-truncated equations of motion apply at all times $t$.

The $\ddot f_1$ in Eq.~(\ref{eqe1}) was restricted to $O(\epsilon^0)$.  From the solution (\ref{om}) one finds that indeed there are subleading corrections
\beq
\ddot f_1(\epsilon x,t)=-\Omega^2f_1(\epsilon x,t)=-\left(m^2+
2\epsilon m\Omega_1+\epsilon^2(\Omega_1^2+2m\Omega_2)\right)f_1(\epsilon x,t)+O(\epsilon^3)
\eeq
which will appear in the equations at higher orders.  

In particular, at order $O(\epsilon^2)$ in (\ref{eomf}), $\Omega_1$ would lead to a resonance.  As this is already well known \cite{fodor}, we will simply set $\Omega_1=0$.  Thus
\beq
\Omega=m+\epsilon^2\Omega_2+O(\epsilon^3) \label{o2eq}.
\eeq
So far, the parameter $\epsilon$ has been arbitrary.  On the other hand, $\Omega$ is a physical parameter, corresponding to an observable property of our solution.  $\Omega_2$, on the other hand, is a function of $\epsilon$ of $\Omega$ given by solving Eq.~(\ref{o2eq}).   Therefore, we are free to transform $\epsilon\rightarrow \epsilon\p=\epsilon\sqrt{-2m\Omega_2}$ and drop the prime, so long as $\Omega_2$ is negative.
We will make this transformation, restricting our attention to solutions with frequencies below $m$.  Whatever dimensions $[\epsilon]$ one assigned to $\epsilon$ before the transformation, $\Omega_2$ had dimensions of $[m/\epsilon^2]$, and so after the transformation $\epsilon$ will have dimensions of mass.

After the transformation, keeping the physical quantity $\Omega$ fixed, we find
\beq
\Omega=\sqrt{m^2-\epsilon^2}+O(\epsilon^3)
\eeq
corresponding to $\Omega_2=-1/(2m)$.

\subsubsection{Order $O(\epsilon^2)$}

At the next order, Eq.~(\ref{eom}) becomes
\bea
\ddot f_2(\epsilon x,t)\Big\vert_{O(\epsilon^0)}+m^2 f_2(\epsilon x,t)&=&-\frac{g V^{(3)}(0)}{2}f_1^2(\epsilon x,t)\\
&=&-\frac{g V^{(3)}(0)}{2}\left[
F^2(\epsilon x)e^{2i\Omega t}+F^{*2}(\epsilon x)e^{-2i\Omega t}+2|F(\epsilon x)|^2
\right].\nonumber
\eea
The homogeneous solution for $f_2$ can be absorbed into an $O(\epsilon)$ contribution to $F$, whereas the particular solution is
\beq
f_2(\epsilon x,t)=\frac{g V^{(3)}(0)}{6m^2}\left[
F^2(\epsilon x)e^{2i\Omega t}+F^{*2}(\epsilon x)e^{-2i\Omega t}-6|F(\epsilon x)|^2
\right].
\eeq

\subsubsection{Order $O(\epsilon^3)$}

At this order, there is a contribution from
\beq
\ddot f_1(\epsilon x,t)\Big\vert_{O(\epsilon^2)}=\epsilon^2 f_1(\epsilon x,t). \label{ddf1}
\eeq
Including the $O(\epsilon^2)$ term in Eq.~(\ref{ddf1}) on the left hand side, one obtains, at the next order
\bea
\ddot f_3(\epsilon x,t)\Big\vert_{O(\epsilon^0)}+m^2 f_3(\epsilon x,t)+f_1&=&\partial^2_{\epsilon x} f_1(\epsilon x,t)\Big\vert_{O(\epsilon^0)}-g V^{(3)}(0)f_1(\epsilon x,t)f_2(\epsilon x,t)\nonumber\\
&&\hspace{-4cm}\ \ -\frac{g^2V^{(4)}(0)}{6}f_1^3(\epsilon x,t)\nonumber\\
&&\hspace{-4cm}=\left[
F\pp(\epsilon x)e^{i\Omega t}+F^{\prime\prime *}(\epsilon x)e^{-i\Omega t}
\right]
-\frac{g^2V^{(4)}(0)}{6}\left[
F(\epsilon x)e^{i\Omega t}+F^{ *}(\epsilon x)e^{-i\Omega t}
\right]^3
\nonumber\\
&&\hspace{-4cm}\ \ +\frac{g^2 V^{(3)2}(0)}{6m^2}\left[
F(\epsilon x)e^{i\Omega t}+F^{ *}(\epsilon x)e^{-i\Omega t}
\right]\nonumber\\
&&\hspace{-4cm}\ \ 
\times\left[
-F^2(\epsilon x)e^{2i\Omega t}-F^{*2}(\epsilon x)e^{-2i\Omega t}+6|F(\epsilon x)|^2
\right].\label{o3}
\eea
In particular, the term proportional to $e^{i\Omega t}$ is
\beq
\ddot f_3(\epsilon x,t)\Big\vert_{O(\epsilon^0)}+m^2 f_3(\epsilon x,t)\supset e^{i\Omega t}\left[
F\pp(\epsilon x)-F(\epsilon x)+\lambda_F|F(\epsilon x)|^2 F(\epsilon x)
\right] \label{pad}
\eeq
where the $\supset$ symbol indicates that we restrict to the $e^{i\Omega t}$-dependent piece and we have defined
\beq
\lambda_F=g^2\left(\frac{5V^{(3)2}(0)}{6m^2}-\frac{V^{(4)}(0)}{2}\right).
\eeq

The condition that the resonance vanishes
\beq
F\pp(\epsilon x)+\lambda_F|F(\epsilon x)|^2 F(\epsilon x)-F(\epsilon x)=0 \label{fpad}
\eeq
is just the master equation \cite{fodor}, which is solved by
\beq
F(\epsilon x)=e^{i\theta}\sqrt\frac{2}{\lambda_F}{\rm{sech}}\left[ 
\epsilon (x-x_0)
\right] \label{fsol}
\eeq
where $\theta$ and $x_0$ are constants of integration.  The $\lambda_F$ here is equal to $4\lambda$ in the notation of Ref.~\cite{fodor}, as $F=p_1/2$.

The term proportional to $e^{3i\Omega t}$ is
\beq
\ddot f_3(\epsilon x,t)\Big\vert_{O(\epsilon^0)}+m^2 f_3(\epsilon x,t)\supset -\frac{e^{3i\Omega t}}{6}F^3(\epsilon x)\left[g^2V^{(4)}(0)+\frac{g^2V^{(3)2}(0)}{m^2}
\right]
\eeq
leading to
\beq
f_3(\epsilon x,t)=\frac{e^{3i\Omega t}F^3(\epsilon x)+e^{-3i\Omega t}F^{3*}(\epsilon x)}{48m^2}\left[g^2V^{(4)}(0)+\frac{g^2V^{(3)2}(0)}{m^2}
\right]
\eeq
where we have included the identical term with a positive frequency

\section{High Energy Perturbations} \label{altsez}

\subsection{The Ansatz}

We will first consider perturbations $\g$ with wave number of order $O(m)$
\beq
\phi(x,t)=f(\epsilon x,t)+\g (x,t)
=\sum_{n=1}^\infty \epsilon^n f_n(\epsilon x,t) + \delta \sum_{n=1}^\infty \epsilon^n \g_{ n}(x,t). \label{exp}
\eeq
Note that  their wavelengths $O(1/m)$ are much shorter than the oscillon wavelength $O(1/(\epsilon m))$, and so one may expect them to be amenable to a WKB expansion.

We will be interested in the limit
\beq
\epsilon\rightarrow 0\hsp
\delta\rightarrow 0\hsp
\frac{\epsilon^{N+1}}{\delta}\rightarrow 0\hsp
\frac{\delta}{\epsilon^{N}}\rightarrow 0
\eeq
for some positive integer $N$.  Since we are interested in the linearized solution for $\g(x)$, terms of order $O(\delta^2)$ will be identified with zero.

Note that, up to order $O(\epsilon^N)$, we can ignore $\delta$ terms and so we can trust the $\epsilon$ expansion above.  On the other hand, at order $O(\delta)$, the equation of motion (\ref{eom}) is
\beq
\ddot\g(x,t)=\partial_x^2\g(x,t)-V\pp(g f(\epsilon x,t))\g(x,t). \label{sl}
\eeq
Inserting
\bea
V\pp(g f(\epsilon x,t))&=&\sum_{n=0}^\infty g^n\frac{V^{(n+2)}(0)}{n!}f^n(\epsilon x,t)\\
&=&m^2+\epsilon gV^{(3)}(0) f_1+\epsilon^2\left[ 
gV^{(3)}(0)f_2(\epsilon x,t)+\frac{g^2V^{(4)}(0)}{2}f^2_1(\epsilon x,t)
\right]+O(\epsilon^3)\nonumber
\eea
our master formula for perturbations becomes
\bea
\ddot\g(x,t)+m^2\g(x,t)&=&\partial_x^2\g(x,t)-\epsilon gV^{(3)}(0) f_1\g(x,t)\label{pe}\\
&&-\epsilon^2\left[ 
gV^{(3)}(0)f_2(\epsilon x,t)+\frac{g^2V^{(4)}(0)}{2}f^2_1(\epsilon x,t)
\right]
\g(x,t)+O(\delta\epsilon^4).\nonumber
\eea


\subsection{Order $O(\delta\epsilon)$}

The leading order equation for the perturbations is at order $O(\delta\epsilon)$
\beq
\ddot\g_1(x,t)\Big\vert_{O(\epsilon^0)}=\partial_x^2\g_1(x,t)\Big\vert_{O(\epsilon^0)}-m^2\g_1(x,t). \label{od}
\eeq
A basis of positive frequency solutions is given by
\bea
\g_{1}(x,t)&=&\hat G(\epsilon, x)e^{-i(\Omega+ \ok{}) t}\hsp
\hat G(\epsilon,x)=G_+(\epsilon,x)e^{ikx}+G_-(\epsilon,x)e^{-ikx}\nonumber
\\ 
\ok{}&=&\sqrt{m^2+k^2}-\Omega+\sum_{n=1}^\infty \epsilon^n\omega_{k,n}. \label{base}
\eea
Note that $\ok{}$ is positive as $\Omega<m$.  For now $G_\pm(\epsilon,x)$ are essentially arbitrary, but their spatial derivatives are suppressed by at least one power of $\epsilon$.
\beq
\partial_x G_\pm(\epsilon,x)\Big\vert_{O(\epsilon^0)}=0.
\eeq

Later we will need the expansion of $\ddot \g_1$ out to order $O(\epsilon^2)$
\bea
\ddot\g_1(x,t)&=&-(\Omega+ \ok{})^2\g_1(x,t)=-\left(
\sqrt{m^2+k^2}+\sum_{n=1}^\infty \epsilon^n\omega_{k,n}\right)^2\g_1(x,t)\\
&=&-\left[(m^2+\ok{}^2)+2\epsilon\sqrt{m^2+k^2}\omega_{k,1}+\epsilon^2\left(\omega^2_{k,1}+2\sqrt{m^2+k^2}\omega_{k,2}
\right)
\right]\g_1(x,t).\nonumber
\eea

\subsection{Order $O(\delta\epsilon^2)$}

At the next order, one finds
\bea
\left[\ddot\g_{2}(x,t)+(m^2-\partial_x^2)\g_{2}(x,t)\right]\Big\vert_{O(\epsilon^0)}&=&\frac{1}{\epsilon}\left(-\ddot\g_{1}(x,t)+\g\pp_{1}(x,t)\right)\Big\vert_{O(\epsilon)}{ - }g V^{(3)}(0)f_1(\epsilon x,t)\g_1(x,t)\nonumber\\
&&\hspace{-4cm}=\left[2\left(\sqrt{m^2+k^2}\omega_{k,1}\hat G(\epsilon, x)
+ik\left(e^{ikx}\partial_{\epsilon x}G_+(\epsilon, x)-e^{-ikx}\partial_{\epsilon x}G_-(\epsilon,x)
\right)\right)e^{-i\Omega t}\right.\nonumber\\
&&\hspace{-3.5cm}\left.
{ - }g V^{(3)}(0)
\left(
F(\epsilon x)+F^*(\epsilon x)e^{-2i\Omega t}
\right)\hat G(\epsilon, x)\right]e^{-i\ok{} t}\label{eq2}
.
\eea
Again the homogeneous solutions for $\g_2$ can be rescaled into the coefficients of the functions $\g_1$, leaving the particular solution.  The $e^{-i(\Omega+\ok{}) t}$ term potentially leads to a resonance.  It is a first order, homogeneous ordinary differential equation whose solution contains one constant of integration, that can be interpreted as $k$.  We will consider the following solution, which is equivalent to any other choice for some $k$,
\beq
\omega_{k,1}=0\hsp \partial_x G_+(\epsilon, x)\Big\vert_{O(\epsilon)}e^{ikx}=\partial_x G_-(\epsilon, x)\Big\vert_{O(\epsilon)}e^{-ikx}.
\eeq

Choose the Ansatz
\beq
\g_{2}(x,t)=\hat G(\epsilon, x)\left(
c_k F(\epsilon x) e^{-i\ok{}t}+d_k F^*(\epsilon x) e^{-i(2\Omega+\ok{})t}\right).
\eeq
Now substitute this into (\ref{eq2}).  Following the usual logic of the expansion of Ref.~\cite{fodor}, we drop $\partial_x(F(\epsilon x)G(\epsilon,x)$ terms as they are of higher order in $\epsilon$.  The coefficient of $F(\epsilon x) \hat G(\epsilon,x)e^{-i\ok{}t}$ is
\beq
\left[-\ok{}^2+m^2+k^2\right]c_k=m(2\ok{}+m)c_k=-gV^{(3)}(0)
\eeq
while the coefficient of $F^*(\epsilon x)\hat G(\epsilon,x) e^{-i(2\Omega+\ok{})t}$ is 
\beq
\left[-(2\Omega+\ok{})^2+m^2+k^2\right]d_k=-m(2\ok{}+3m)d_k=-gV^{(3)}(0).
\eeq
In summary, we find
\bea
\g_{2}(x,t)&=&gV^{(3)}(0)\hat G(\epsilon, x) e^{-i\omega_k t}\left[-
\frac{F(\epsilon x)}{m(m+2\ok{})}+\frac{F^*(\epsilon x)}{m(3m+2\ok{})} e^{-2i\Omega t}\right].
\eea

\subsection{Order $O(\delta\epsilon^3)$}

The next order equation is
\bea
\ddot \g_3(x,t)\Big\vert_{O(\epsilon^0)}+m^2 \g_3(x,t)-\g\pp_3(x,t)&=&\frac{\g\pp_1(x,t)}{\epsilon^2}\Big\vert_{O(\epsilon^2)}+\frac{\g\pp_2(x,t)}{\epsilon}\Big\vert_{O(\epsilon)}\nonumber\\
&&\hspace{-3cm}
-\frac{g^2V^{(4)}(0)}{2}f_1^2(\epsilon x,t)\g_1(x,t)-\frac{\ddot\g_1(x,t)}{\epsilon^2}\Big\vert_{O(\epsilon^2)}\nonumber\\
&&\hspace{-3cm}-{g V^{(3)}(0)} \left(f_1(\epsilon x,t)\g_2(x,t)+f_2(\epsilon x,t)\g_1(x,t)
\right) \label{cort3}
\eea
where
\bea
-\ddot\g_1(x,t)\Big\vert_{O(\epsilon^2)}&=&2\sqrt{m^2+k^2}\omega_{k,2}\hat G(\epsilon, x)e^{-i(\Omega+\omega_k) t}
\\
\g\pp_{1}(x,t)\Big\vert_{O(\epsilon^2)}&=&
\left[e^{ikx}\left(2ik
\partial_x+\partial^2_x\right) G_+(\epsilon, x)\Big\vert_{O(\epsilon^2)}\right.\nonumber\\
&&\left.
+
e^{-ikx}\left(-2ik
\partial_x+\partial^2_x\right) G_-(\epsilon, x)\Big\vert_{O(\epsilon^2)}
\right]e^{-i(\Omega+\omega_k) t}\nonumber
\\
\g\pp_{2}(x,t)\Big\vert_{O(\epsilon)}&=&2igkV^{(3)}(0) \left( G_-(\epsilon, x)e^{-ikx}-G_+(\epsilon, x)e^{ikx}\right)e^{-i\omega_k t}\nonumber\\
&&\times\left[
\frac{F\p(\epsilon x)}{m(m+2\ok{})}-\frac{F^{\prime *}(\epsilon x)}{m(3m+2\ok{})} e^{-2i\Omega t}\right]\nonumber
\eea
and
\bea
-\frac{g^2V^{(4)}(0)}{2}f_1^2(\epsilon x,t)\g_1(x,t)&=&-\frac{g^2V^{(4)}(0)}{2}\left[F(\epsilon x)e^{i\Omega t}+F^*(\epsilon x)e^{-i\Omega t}
\right]^2
\hat G(\epsilon, x)e^{-i(\Omega+\ok{}) t}\nonumber\\
-{g V^{(3)}(0)} f_1(\epsilon x,t)\g_2(x,t)
&=&{g^2 V^{(3)2}(0)}\left[\frac{F^2(\epsilon x)}{m(m+2\ok{})} e^{i\Omega t} -\frac{F^{2*}(\epsilon x)}{m(3m+2\ok{})} e^{-3i\Omega t}\right.\nonumber\\
&&\left.+\frac{2m|F(\epsilon x)|^2}{m(4\ok{}^2+8m\ok{}+3m^2)}e^{-i\Omega t}\right]\hat G(\epsilon, x)e^{-i\omega_k t}\nonumber\\
{g V^{(3)}(0)} f_2(\epsilon x,t)\g_1(x,t)&=&\frac{g^2 V^{(3)2}(0)}{6m^2} \left[6|F(\epsilon x)|^2e^{-i\Omega t}-
F^2(\epsilon x)e^{i\Omega t}\right.\\
&&\left.-F^{*2}(\epsilon x)e^{-3i\Omega t}
\right]
\hat G(\epsilon, x)e^{-i\omega_k t}.\nonumber
\eea

Again, at this order the $x$ dependence of $\g_3(x,t)$ is that of the right hand side of (\ref{cort3}).  The derivatives in $\g_3\pp$, at this order, only act on the $e^{\pm ikx}$ and so the left hand side is $\ddot\g_3+(\Omega+\ok{})^2\g_3$.  This means that terms on the right hand side with frequency $\Omega+\ok{}$ lead to a resonance.  These terms are
\bea
\ddot\g_3+\ok{}^2\g_3&\supset& e^{-i(\Omega+\ok{})t}\left[ 
2\sqrt{m^2+k^2}\omega_{k,2}\hat G(\epsilon, x)
+C_k|F(\epsilon x)|^2\hat G(\epsilon, x)\right.\nonumber\\
&&\left.
+
e^{ikx}\left(2ik
\partial_x+\partial^2_x\right) G_+(\epsilon, x)\Big\vert_{O(\epsilon^2)}
+
e^{-ikx}\left(-2ik
\partial_x+\partial^2_x\right) G_-(\epsilon, x)\Big\vert_{O(\epsilon^2)}
\right]\nonumber \\
C_k&=&-g^2V^{(4)}(0)+\frac{4\ok{}^2+8m\ok{}+5m^2}{m^2(4\ok{}^2+8m\ok{}+3m^2)}g^2V^{(3)2}(0).\label{altse}
\eea
The resonance is removed by setting the term in the square bracket to zero
\bea
e^{ikx}\left(2ik
\partial_x+\partial^2_x\right) G_+(\epsilon, x)\Big\vert_{O(\epsilon^2)}
&+&
e^{-ikx}\left(-2ik
\partial_x+\partial^2_x\right) G_-(\epsilon, x)\Big\vert_{O(\epsilon^2)}\nonumber\\
&=&-\left(2\sqrt{m^2+k^2}\omega_{k,2}
+C_k|F(\epsilon x)|^2\right)\hat G(\epsilon, x).\label{nores3}
\eea
This fixes $\hat G(\epsilon, x)$ to be the solution of this nonrelativistic Schrodinger equation, albeit with a $k$-dependent potential, and leading to the usual small wave number and amplitude shift found for a high frequency wave in a wide, shallow potential~$-C_k|F|^2/2$.  As the potential depends on $k$, or more precisely $\ok{}/m$, one expects that the solutions will not be orthogonal when the perturbation is relativistic, corresponding to $k$ of order $m$ or greater.  


Now one can easily solve for $\g_3$
\bea
\g_3(x,t)&=&\frac{F^2(\epsilon x)\hat G(\epsilon, x)}{4m\ok{}}e^{i(\Omega-\ok{})t}\left[-\frac{g^2V^{(4)}(0)}{2}
+g^2 V^{(3)2}(0)\left(-\frac{1}{6m^2}+\frac{1}{m(m+2\ok{})}
\right)\right]\nonumber\\
&&\hspace{-1cm}+\frac{F^{*2}(\epsilon x)\hat G(\epsilon, x)}{4m(2m+\ok{})}e^{-i(3\Omega+\ok{})t}\left[\frac{g^2V^{(4)}(0)}{2}
+g^2 V^{(3)2}(0)\left(\frac{1}{6m^2}+\frac{1}{m(3m+2\ok{})}
\right)\right]\nonumber\\
&&\hspace{-1cm}+\frac{2igkV^{(3)}(0)\left( G_-(\epsilon, x)e^{-ikx}-G_+(\epsilon, x)e^{
ikx}\right)}{m^2}\nonumber\\
&&\times\left[\frac{F^{\prime}(\epsilon x)}{(2\ok{}+m)^2 }e^{-i\ok{}t}+\frac{F^{\prime *}(\epsilon x)}{(2\ok{}+3m)^2 }e^{-i(2\Omega+\ok{})t}\right].
\eea

In the nonrelativistic limit, $\ok{}=0$, we find $C_k=2\lambda_F$.   In particular, the $k$-dependence of the potential disappears, as is expected for a nonrelativistic particle.

\subsection{Solving the No-Resonance Condition}
The second spatial derivative yields terms at two different orders
\beq
\partial^2_{\epsilon x}\hat G(\epsilon,x)=\frac{1}{\epsilon^2}\partial^2_{x}\hat G(\epsilon,x)=\frac{e^{ikx}}{\epsilon^2}(-k^2+2ik\partial_x+\partial^2_x)G_+(\epsilon, x)+\frac{e^{-ikx}}{\epsilon^2}(-k^2-2ik\partial_x+\partial^2_x)G_-(\epsilon, x)
\eeq
where the first equality is the definition of the differential operator $\partial^2_{\epsilon x}$.  One recognizes the last two terms from (\ref{altse})
\beq
e^{ikx}(2ik\partial_x+\partial^2_x)G_+(\epsilon, x)+e^{-ikx}(-2ik\partial_x+\partial^2_x)G_-(\epsilon, x)=k^2\hat G(\epsilon,x)+\epsilon^2 \partial^2_{\epsilon x}\hat G(\epsilon,x). \label{sub}
\eeq

Substituting Eq.~(\ref{sub}) into the no-resonance condition Eq.~(\ref{nores3}) one finds
\bea
-\left(\frac{k^2}{\epsilon^2}+2\sqrt{m^2+k^2}\omega_{k,2}\right)\hat G(\epsilon,x)&=&\partial_{\epsilon x}^2\hat G(\epsilon,x)+C_k |F(\epsilon x)|^2\hat G(\epsilon,x)\label{pt1}\\
&=&\partial_{\epsilon x}^2\hat G(\epsilon,x)+ \frac{2C_k}{\lambda_F} \sech^2\left(\epsilon(x-x_0)\right)\hat G(\epsilon,x).\nonumber
\eea
This is of the P\"oschl-Teller form, although we remind the reader that for each wave number $k$ one must choose the potential $C_k$.  

This equation is of the general P\"oschl-Teller form
\beq
z\pp(\hat x)+(D\sech^2(\hat x)+E)z(\hat x)=0\hsp
D=\frac{2C_k}{\lambda_F}\hsp
E=\frac{k^2}{\epsilon^2}+2\sqrt{m^2+k^2}\omega_{k,2}\hsp\hat x=\epsilon(x-x_0). \label{pte}
\eeq
Define
\beq
y(\hat x)=\cosh^2(\hat x)\hsp Z(y)=z(\hat x)=\hat G(\epsilon,x).
\eeq
Then
\beq
\frac{\partial y(\hat x)}{\partial \hat x}
=2\sqrt{y(y-1)}\hsp \frac{\partial \sqrt{y(\hat x)}}{\partial \hat x}=\sqrt{y-1}\hsp \frac{\partial \sqrt{y(\hat x)-1}}{\partial \hat x}=\sqrt{y}\hsp \frac{\partial ^2y(\hat x)}{\partial \hat x^2}=4y-2
\eeq
and so
\bea
\frac{\partial Z(y(\hat x))}{\partial \hat x}&=&\frac{\partial y(\hat x)}{\partial \hat x} Z\p(y(\hat x))=2\sqrt{y(y-1)} Z\p(y)\\
\frac{\partial^2 Z(y(\hat x))}{\partial \hat x^2}&=&\left(\frac{\partial y(\hat x)}{\partial \hat x}\right)^2 Z\pp(y(\hat x))+\frac{\partial^2 y(\hat x)}{\partial\hat x^2} Z\p(y(\hat x))\nonumber\\
&=&4y(y-1)Z\pp(y)+(4y-2)Z\p(y).\nonumber
\eea

To obtain the hypergeometric equation, we need to multiply $Z$ by $y^c$ for some $k$-dependent $c$
\beq
z(\hat x)=Z(y(\hat x))=y^c(\hat x)\hat Z(y(\hat x)).
\eeq
Using
\beq
\frac{\partial y^c(\hat x)}{\partial \hat x}=cy^{c-1}\frac{\partial y(\hat x)}{\partial \hat x}=2cy^{c-1}\sqrt{y(y-1)}\hsp
\frac{\partial^2 y^c(\hat x)}{\partial \hat x^2}
=4c^2y^c+c(2-4c)y^{c-1}
\eeq
one finds
\bea
z\pp(\hat x)&=&\frac{\partial^2 y^c(\hat x)}{\partial \hat x^2}\hat{Z}(y(\hat x))+2\frac{\partial y^c(\hat x)}{\partial \hat x }\frac{\partial \hat Z(y(\hat x))}{\partial \hat x}+y^c(\hat x)\frac{\partial^2 \hat Z(y(\hat x))}{\partial \hat x^2}\\
&=&y^c\left[c\left(4c+\frac{2-4c}{y}\right)\hat{Z}(y) 
+\frac{2c}{y}2\sqrt{y(y-1)}2\sqrt{y(y-1)}\hat Z\p(y)\right.\nonumber\\
&&\ \ \ \left. 4y(y-1)\hat Z\pp(y)+(4y-2)\hat Z\p(y)
\right]\nonumber\\
&=&4y^c\left[ y(y-1) \hat Z\pp(y) +\left(\left(1+2c\right)y-2c-\frac{1}{2}\right)\hat Z\p(y)+c\left(c+\frac{1/2-c}{y}\right)\hat{Z}(y)
\right]\nonumber\nonumber\\
&=&4y^c \left(-\frac{D/4}{y}-\frac{E}{4}
\right)\hat{Z}(y)\nonumber
\eea
where the last equality used the P\"oschl-Teller equation Eq.~(\ref{pte}).  Fixing
\beq
c\left(4c-2\right)=D
\eeq
so that
\beq
c=\frac{1\pm\sqrt{1+4D}}{4} \label{ceq}
\eeq
the $\hat Z/y$ terms cancel in the last equality.  This leaves the hypergeometric equation
\beq
y(1-y) \hat Z\pp(y) +\left(2c+\frac{1}{2}-\left(2c+1\right)y\right)\hat Z\p(y)-\left(c^2+\frac{E}{4}\right)\hat{Z}(y)=0
\eeq
which has coefficients
\beq
C=2c+\frac{1}{2}\hsp AB=c^2+\frac{E}{4}\hsp A+B=2c.
\eeq
One finds
\beq
0=A^2-{2cA}+c^2+\frac{E}{4}\hsp A=c\pm\frac{i\sqrt{E}}{2}\hsp B=c\mp\frac{i\sqrt{E}}{2}.
\eeq
The independent solutions are then 
\bea
z(x)&=&\cosh^{2c}(\epsilon (x-x_0)) {}_2F_1\left(A,B;1+A+B-C;-\sinh^2(\epsilon (x-x_0))\right)\\
&=&\sqrt{\cosh(\epsilon (x-x_0))}^{{1\pm\sqrt{1+4D}}}\nonumber\\
&&\times {}_2F_1\left(c\pm\frac{i\sqrt{E}}{2},c\mp\frac{i\sqrt{E}}{2};\frac{1}{2};-\sinh^2(\epsilon (x-x_0))\right)\nonumber\\
z(x)&=&\cosh^{2c}(\epsilon (x-x_0))(-\sinh^2(\epsilon(x-x_0))^{C-A-B} \nonumber\\
&&\times{}_2F_1\left(C-A,C-B;1+C-A-B;-\sinh^2(\epsilon (x-x_0))\right)\nonumber\\
&=&i\ \sinh(\epsilon (x-x_0)) \sqrt{\cosh(\epsilon (x-x_0))}^{{1\pm\sqrt{1+4D}}}\nonumber\\
&&\times{}_2F_1\left(\frac{1}{2}+c\mp\frac{i\sqrt{E}}{2},\frac{1}{2}+c\pm\frac{i\sqrt{E}}{2};\frac{3}{2};-\sinh^2(\epsilon (x-x_0))\right)\nonumber
\eea
where the sign in the argument of $F$ is independent of the sign in the exponent.  
The sign in the argument, on the other hand, is related to the direction of propagation of the wave. 

We remind the reader that
\beq
1+4D=1+\frac{8C_k}{\lambda_F}\hsp 1-4c^2=\frac{1-2D\mp \sqrt{1+4D}}{2}=\frac{1-4C_k/\lambda_F\mp \sqrt{1+8C_k/\lambda_F}}{2}
\eeq
Here $C_k/\lambda_F$ is a relativistic correction.  For long wavelength perturbations it equals $2$, but as the wave number approaches the meson mass, it becomes dependent on the individual couplings and it decreases.

The solutions in this regime have a continuous spectrum, and so any $\omega_{k,2}$ of order $O(\epsilon^0)$ corresponds to two solutions.  We will fix a unique pair of solutions at each $k$ by setting $\omega_{k,2}=0$.  Then, the term $\sqrt{E}$ is equal to
\beq
\sqrt{E}=k/\epsilon.
\eeq

One may simply replace each $\pm\sqrt{E}$ with $\pm k/\epsilon$ and then check that the asymptotic wavenumber is indeed $k$.  Recall that only the asymptotic wavenumber $k$ solution, where $k$ agrees with the coefficient of $C_k$, solves the no resonance condition.  With this substitution, the solutions are
\bea
\hat G(\epsilon,x)&=&\sqrt{\cosh(\epsilon (x-x_0))}^{{1\pm\sqrt{1+8C_k/\Lambda_F}}}\\
&&\times {}_2F_1\left(\frac{1\pm\sqrt{1+8C_k/\lambda_F}}{4} \pm\frac{ik}{2\epsilon},\frac{1\pm\sqrt{1+8C_k/\lambda_F}}{4} \mp\frac{ik}{2\epsilon};\frac{1}{2};-\sinh^2(\epsilon (x-x_0))\right)\nonumber\\
\hat G(\epsilon,x)&=&i\ \sinh(\epsilon (x-x_0)) \sqrt{\cosh(\epsilon (x-x_0))}^{{1\pm\sqrt{1+8C_k/\Lambda_F}}}\nonumber\\
&&\times{}_2F_1\left(\frac{3\pm\sqrt{1+8C_k/\lambda_F}}{4} \mp\frac{ik}{2\epsilon},\frac{3\pm\sqrt{1+8C_k/\lambda_F}}{4} \pm\frac{ik}{2\epsilon};\frac{3}{2};-\sinh^2(\epsilon (x-x_0))\right)\nonumber
\eea
Using the plus sign, these are the even and odd normal modes of the $\phi^4$ kink when $C_k=3\lambda_F$ and the Sine-Gordon soliton when $C_k=\lambda_F$.  
Note that $\lambda_F>0$ if an oscillon exists and also $\ok{}>0$.  Therefore 
\beq
-g^2V^{(4)}(0)+\frac{g^2V^{(3)2}(0)}{m^2} < C_k
=-g^2V^{(4)}(0)+\left(
1+\frac{2m^2}{4\ok{}^2+8m\ok{}+3m^2}\right)\frac{g^2V^{(3)2}(0)}{m^2}\leq 2\lambda_F
\eeq
implies that $C_k$ is monotonically decreasing with respect to $\ok{}$.  In particular it is greatest at $\ok{}=0$, where $C_k=2\lambda_F$, and so it is never equal to $3\lambda_F$.  

We can read the reflection and transmission coefficients, as well as the phase shift, off of these functions.  For example, the norm squared of the reflection coefficient $R$ is given by
\beq
|R|^2=\cos^2 (\phi_e-\phi_o)
\eeq
where $\phi_e$  and $\phi_o$ are the asymptotic phases of the even and odd solutions.  For the P\"oschl-Teller problem with 
\beq
\lambda_{PT}(\lambda_{PT}-1)=\frac{2C_k}{\lambda_F}.
\eeq
According to Eq.~(39.21) of Ref.~\cite{flugge}, this phase difference is
\beq
\phi_e-\phi_o={\rm{arctan}}\left(\frac{\sinh(\pi k/\epsilon)}{\rm{sin}(\pi\lambda_{PT})}
\right).
\eeq
As a result
\beq
|R|^2=\frac{\rm{sin}^2(\pi\lambda_{PT})}{\rm{sin}^2(\pi\lambda_{PT})+{\sinh^2(\pi k/\epsilon)}}.
\eeq
The treatment in this section is only valid when $k>>\epsilon$, a condition which is relaxed in Sec.~\ref{bassez}.  If $k$ is, for example, of order $O(\epsilon^0)$, then $|R|^2$ is of order $O(e^{-1/\epsilon})$ and so vanishes in the $\epsilon$ expansion.  More generally, we learn that oscillons do not reflect small-amplitude radiation of wavelength much less than the oscillon width $1/\epsilon$. 

Theories with $V^{(3)}(0)=0$, like the Sine-Gordon model, are much simpler.  Then $C_k=2\lambda_F$ for all $k$.  In particular, the normal modes with wave number $k\gg\epsilon m$ are solutions of the same P\"oschl-Teller problem, and so are orthogonal.  It would be interesting to understand how this property relates to the stability of the Sine-Gordon breather.

\subsection{Longer Wavelength Perturbations}

But what about zero modes and other bound modes?  These have compact support and so are not, at leading order, approximated by plane waves $\g_1$.  Why were they missed in the treatment above?  We assumed that perturbations had wave numbers of order $O(m \epsilon^0)$ in the Ansatz (\ref{exp}).  As the only mass scales in the problem are the meson mass $m$ and the inverse oscillon length $\epsilon m$, if the wave number is much larger than $O(m \epsilon^0)$ it is insensitive to the oscillon and to the mass and so is as in the free theory.  On the other hand, a smaller wave number invalidates the $\epsilon$ expansion above.

To address such long wavelengths, in the following two sections, we rename the perturbation $\g(\epsilon^{N/2} x,t)$ when $N=1$ and $N=2$ in Secs.~\ref{intsez} and \ref{bassez} respectively.  This describes fluctuations with wavenumber of order $\epsilon^{N/2} m$.  Then the equation of motion (\ref{sl}) becomes
\beq
\ddot\g(\epsilon^{N/2}x,t)=\epsilon^N\partial_{\epsilon^{N/2}x}^2\g(\epsilon^{N/2}x,t)-V\pp(g f(\epsilon x,t))\g(\epsilon^{N/2}x,t).
\eeq

\section{Intermediate Wave Number Perturbations} \label{intsez}

The derivation above fails already for a perturbation with wave number $\sqrt{\epsilon} m$, as the kinetic energy term $(\partial\phi)^2$ enters only at order $O(\delta\epsilon)$.  

More precisely, consider the Ansatz 
\beq
\phi(x,t)=\sum_{n=1}^\infty \epsilon^n f_n(\epsilon x,t) + \delta \sum_{n=1}^\infty \epsilon^n \g_{ n}(\sqrt\epsilon x,t). 
\eeq
Then at order $O(\delta\epsilon)$ instead of (\ref{od}) one finds
\beq
\ddot\g_1(\sqrt\epsilon x,t)\oe{0}=-m^2\g_1(\sqrt\epsilon x,t).\label{into1}
\eeq
This is solved by 
\beq
\g_1(\sqrt\epsilon x,t)=G_k(\sqrt\epsilon x)e^{-i (\Omega+\omega_k) t}\hsp \omega_k=\sum_{n=1}^\infty\epsilon^n\omega_{k,n}
\eeq
for some function $G_k$.  Here $k$ labels the solution.  In particular, up to order $O(\epsilon^2)$
\beq
\ddot\g_1(\sqrt\epsilon x,t)=-(\Omega+\omega_k)^2\g_1(\sqrt\epsilon x,t)=\left(-m^2-2\epsilon m\omega_{k,1}+\epsilon^2(1-\omega_{k,1}^2-2m\omega_{k,2}
)
\right)\g_1(\sqrt\epsilon x,t).
\eeq

Next, at order $O(\delta\epsilon^2)$, Eq.~(\ref{eom}) becomes 
\bea
\ddot\g_2(\sqrt\epsilon x,t)\oe{0}+m^2\g_2(\sqrt\epsilon x,t)&=& 2  m\omega_{k,1}         \g_1(\sqrt\epsilon x,t)+ \partial_{\sqrt\epsilon x}^2\g_1(\sqrt\epsilon x,t)\oe{0}\nonumber\\
&&\hspace{-4cm}\ \ { - } gV^{(3)}(0)f_1(\epsilon x,t)\g_1(\sqrt\epsilon x,t)\nonumber\\
&&\hspace{-4cm}=\left(G_k^{\prime\prime}(\sqrt\epsilon x)+2 \omega_{k,1} mG_k(\sqrt\epsilon x)\right)e^{-i(\Omega+\omega_k) t}\nonumber\\
&&\hspace{-4cm}\ \ \ { - } gV^{(3)}(0)\left[ F(\epsilon x)e^{-i\omega_k t}+F^*(\epsilon x)e^{-i(2\Omega+\omega_k) t}
\right]G_k(\sqrt\epsilon x)
.\label{int2}
\eea
The $G\pp$ terms lead to a naive resonance unless
\beq
G_k^{\prime\prime}(\sqrt\epsilon x)+2  m\omega_{k,1}G_k(\sqrt\epsilon x)=0.
\eeq  
As a result, a basis of solutions is
\beq
G_k(\sqrt\epsilon x)=e^{-i\sqrt{\epsilon}kx}+O(\epsilon)\hsp
k^2=2m\omega_{k,1}. \label{intgk}
\eeq
We see that $\g_1$ is an ordinary plane wave solution, as in the high energy perturbations.  This is to be expected, as the wavelength of the perturbations is still much less than the length of the oscillon.  Note that $k$ is of order $O(\epsilon^0)$ as $m$ and $\omega_{k,1}$ are both of order $O(\epsilon^0)$ by definition.

Summarizing, we have found
\bea
\g_1(\sqrt\epsilon x,t)&=&e^{-i\sqrt\epsilon  kx}
e^{-i(\Omega+\omega_k) t}\\
\g_2(\sqrt\epsilon x,t)&=& \frac{gV^{(3)}(0)}{3m^2}\left[ -3F(\epsilon x)e^{-i\omega_k t}+F^*(\epsilon x)e^{-i(2\Omega+\omega_k) t}
\right]e^{-i\sqrt{\epsilon}kx}+H_k(\sqrt\epsilon x,t)\nonumber
\eea
where $H_k(\sqrt\epsilon x,t)$ is a homogeneous solution.  We will see shortly that the only part of the homogeneous solution needed to remove the leading resonance is
\beq
H_k(\sqrt\epsilon x,t)=H_k(\sqrt\epsilon x)e^{i(\Omega-\omega_k)t}+\mathfrak{G}_k(\epsilon,x)e^{-i(\Omega+\omega_k)t}.
\eeq
One may choose to interpret $\mathfrak{G}_k$ as the $O(\epsilon)$ correction to $G_k$ in Eq.~(\ref{intgk})
\beq
G_k(\sqrt\epsilon x)=e^{-i\sqrt{\epsilon}kx}+\epsilon \mathfrak{G}_k(\epsilon,x)+O(\epsilon^2). \label{gg}
\eeq

Let us proceed to order $O(\delta\epsilon^3)$, where we have seen that $F^2G$ resonances can appear.  Now the equation of motion is
\bea
\ddot \g_3(\sqrt\epsilon x,t)\oe{0}+m^2 \g_3(\sqrt\epsilon x,t)&=&k^2\frac{gV^{(3)}(0)}{3m^2}\left[ 3F(\epsilon x)e^{-i\omega_k t}+F^{*}(\epsilon x)e^{-i(2\Omega+\omega_k) t}
\right]e^{-i\sqrt{\epsilon}kx}
\nonumber\\
&&\hspace{-2cm}
-k^2 H_k(\sqrt\epsilon x)e^{i(\Omega-\omega_k)t}+H_k\pp(\sqrt\epsilon x)e^{i(\Omega-\omega_k)t}
\nonumber\\
&&\hspace{-2cm}
+k^2 \mathfrak{G}_k(\epsilon, x)e^{i(\Omega-\omega_k)t}+\partial^2_{\sqrt{\epsilon}x}\mathfrak{G}_k(\epsilon, x)e^{i(\Omega-\omega_k)t}\oe{0}
\nonumber\\
&&\hspace{-2cm}+\epsilon^2 \left(\omega^2_{k,1}+2m\omega_{k,2}-1\right)\g_1(\sqrt\epsilon x,t)
-\frac{g^2V^{(4)}(0)}{2}f_1^2(\epsilon x,t)\g_1(\sqrt\epsilon x,t)\nonumber\\
&&\hspace{-2cm}-{g V^{(3)}(0)} \left(f_1(\epsilon x,t)\g_2(\sqrt\epsilon x,t)+f_2(\epsilon x,t)\g_1(\sqrt\epsilon x,t)
\right).\label{int3}
\eea

Resonances can arise both from terms proportional to $e^{i(\Omega-\omega_k) t}$ and also those proportional to $e^{-i(\Omega+\omega_k) t}$.  The terms proportional to $e^{i(\Omega-\omega_k) t}$ are
\beq
\ddot \g_3(\sqrt\epsilon x,t)\oe{0}+m^2 \g_3(\sqrt\epsilon x,t)\supset e^{i(\Omega-\omega_k)t}\left(\lambda_FF^2(\epsilon x)e^{-i\sqrt\epsilon  kx}-k^2 H_k(\sqrt\epsilon x)+H_k\pp(\sqrt\epsilon x)\right). \label{int3}
\eeq

The resonance is therefore canceled if
\beq
H_k\pp(\sqrt\epsilon x)-k^2 H_k(\sqrt\epsilon x)=-\lambda_F|F|^2(\epsilon x)e^{-i\sqrt\epsilon  kx}.
\eeq
This is easily solved by Fourier transform.  The Fourier transform of this equation is
\beq
-\frac{(p^2+\epsilon k^2)}{{\epsilon}}\tilde H_k(p/\sqrt{\epsilon})=-2e^{-ipx_0}\frac{\pi}{\epsilon}(p+\sqrt{\epsilon}k)\csch\left(\frac{\pi(p+\sqrt{\epsilon}k)}{2\epsilon}\right)
.
\eeq
This leads to
\beq
H_k(\sqrt\epsilon x)=2\pi\pin{p} e^{ip(x-x_0)}\frac{p+\sqrt\epsilon k}{p^2+\epsilon k^2}\csch\left(\frac{\pi(p+\sqrt{\epsilon}k)}{2\epsilon}\right). \label{sop}
\eeq
Now recall that $k$ is of order $O(\epsilon^0)$. Therefore the $k/\sqrt\epsilon$ in the argument of csch is of order $O(\epsilon^{-1/2})$ unless $p=-\sqrt{\epsilon}k+O(\epsilon)$.  Note that the pole in $\csch$ at $p=-\sqrt{\epsilon}k$ is canceled by the numerator of the previous term.  As the $\csch$ term leads to an exponential suppression when $p$ is outside of window whose size is of order $\epsilon$, the $p$ integration is over a domain of size $O(\epsilon)$ and so leads to a factor of order $O(\epsilon)$.  As a result, $H_k$ itself is of order $O(\epsilon)$.  However, by definition $H_k$ is a term in $\g_2$ which is of order $O(\epsilon^0)$, and so its order $O(\epsilon)$ corrections are meaningless.  One could define them to be the homogeneous part of $\g_3$, but these will be constrained by the resonance condition at $O(\delta\epsilon^4)$, which we do not consider.  Thus, we set $H_k=0$.


This is to be expected, as the resonance is caused by
\beq
\lambda_FF^2(\epsilon x)e^{-i\sqrt\epsilon  kx}
\eeq
but the characteristic length scale of $F$ is $1/\epsilon$ and that of the phase term is $1/\sqrt\epsilon$.  This means that the phase oscillates of order $O(\epsilon^{-1/2})$ times over the length of the oscillon, leading to destructive interference which already essentially eliminates this potential resonance.  There would not be so much interference if $k$ were of order $O(\sqrt{\epsilon})$, in which case the perturbation length scale would the same as that of $F$.  We next turn to that case in Sec.~\ref{bassez}.  For now, as the $H$ term is exponentially suppressed we will ignore it.

The other resonance corresponds to terms proportional to $e^{-i(\Omega+\omega_k) t}$.  These are
\bea
\ddot \g_3(\sqrt\epsilon x,t)+m^2 \g_3(\sqrt\epsilon x,t)&\supset& e^{-i(\Omega+\omega_k)t}\left[e^{-i\sqrt{\epsilon}kx}\left(
\epsilon^2 \left(\omega^2_{k,1}+2m\omega_{k,2}-1\right)+2\lambda_F|F(\epsilon x,t)|^2\right)\nonumber\right.\\
&&\left.+k^2 \mathfrak{G}_k(\epsilon, x)+\partial^2_{\sqrt{\epsilon}x}\mathfrak{G}_k(\epsilon, x)
\right]. 
\eea
The vanishing of the square brackets is again a linear ordinary differential equation for $\mathfrak{G}_k$ which can be solved by Fourier transform.   Rewriting this as a homogeneous equation for $G_k$ in the form (\ref{gg}), one can identify this as the $O(\epsilon)$ part of a time-independent, non-relativistic Schrodinger equation for a monochromatic wave in a potential well $-\lambda_F|F|^2$
\bea
\left[k^2+\epsilon\left(-1+\frac{k^4}{4m^2}+2m\omega_{k,2}
\right)\right]G_k&=&\partial^2_{\sqrt\epsilon x^2}G_k+2\lambda_F|F(\epsilon x,t)|^2 G_k\nonumber\\
&=& \partial^2_{\sqrt\epsilon x^2}G_k+4\sech^2(\epsilon(x-x_0))G_k. \label{intse}
\eea
Note that this agrees with the nonrelativistic limit of Eq.~(\ref{pt1}), with $G_k$ here identified with $\hat G$ there.   We conclude that intermediate wavelength perturbations are described by the same formulas as the short wavelength perturbations found in the previous section, which is of little surprise as both have wavelengths which are much shorter than the oscillon.


\section{Long Wavelength Perturbations} \label{bassez}

\subsection{Zero Modes}

To find honest, bound modes one needs a wave number which is still lower.  The analysis above fails once the wave number falls to order $O(\epsilon).$  However, if $\g$ depends on position via the combination $\epsilon x$ and it has the same frequency $\Omega$ as the oscillon $f$, then its dependence matches that of $f$ in Eq.~(\ref{exp}).  Therefore, any frequency $\Omega$ solution in this regime, can be obtained by setting $\g=0$ and absorbing it into $f(\epsilon x,t).$  In other words, all such perturbations of wavelength $1/\epsilon$ are already contained in the analysis of Subsec.~\ref{fodor}.  In particular, any solution plus such periodic perturbations should correspond to a solution of the master equation Eq.~(\ref{fpad}).

The solutions of Eq.~(\ref{fpad}) are
\beq
F(\epsilon x)=e^{i\theta}\sqrt\frac{2}{\lambda_F}{\rm{sech}}\left[ 
\epsilon (x-x_0)
\right]. \label{fsol}
\eeq
These depend on two constants of integration.  The first is $x_0$, the position of the center of the oscillon.  The second is the phase $\theta$.  The analogues of zero modes for the oscillon are just the small perturbations in these constants of integration.  In other words, up to normalization constants that may be fixed at will, spatial translations and time translations correspond to the zero modes
\beq
\g_B(\epsilon x,t)=-\partial_{\epsilon x}f(\epsilon x,t)\hsp
\g_T(\epsilon x,t)=- \partial_t f(\epsilon x,t)
\eeq
where $f$ is the sum of all $\epsilon^nf_n$ up to the desired order.  

Note that the term {\it{zero mode}} is rather misleading, as these have frequencies of 
$\Omega$, which is approximately $m$, not zero.  Indeed they are a long wavelength limit of the meson in which the meson no longer fits entirely inside of the oscillon.  In particular, it may be that there is no gap between these zero modes and the continuum.

At leading order $f$ is simply $\epsilon f_1$ where
\beq
f_1(\epsilon x,t)=2\sqrt\frac{2}{\lambda_F}{\rm{sech}}\left[ 
\epsilon (x-x_0)
\right]\cos\left( \Omega t+\theta\right).
\eeq
Therefore the zero modes are
\bea
\g_B(\epsilon x,t)&=& 2\sqrt\frac{2}{\lambda_F}{\rm{tanh}}\left[ 
\epsilon (x-x_0)
\right] {\rm{sech}}\left[ 
\epsilon (x-x_0)
\right]\cos\left( \Omega t+\theta\right)\nonumber\\
\g_T(\epsilon x,t)&=&2m\sqrt\frac{2}{\lambda_F}{\rm{sech}}\left[ 
\epsilon (x-x_0)
\right]\sin\left( \Omega t+\theta\right).
\eea


\subsection{General Long Wavelength Perturbations}

We have discovered that there are two perturbation modes, $\g_B$ and $\g_T$, which have the same frequency $\Omega=\sqrt{m^2-\epsilon^2}$ as the oscillon.  They have compact support, and frequencies below the threshold $m$, and so presumably they are discrete states.  However, our analysis of the continuum modes was only valid for wavenumbers $k$ with $|k|\gg \epsilon$, and so we have not shown that the frequencies of the continuum modes do not extend down to $\Omega$.

Recall that in the case of modes with wavenumbers of order $O(\epsilon)$, the kinetic term, and so the nontrivial constraints, only appear at order $O(\epsilon^3)$.  Here, the $O(\epsilon^2)$ corrections to the frequency also appear.  Thus our above treatment of the $\epsilon$ expansion at $O(\epsilon^3)$ does not apply when the wavenumber is of order $O(\epsilon)$.

Such perturbations need to be considered if one wishes to see whether the set of $\g$ indeed forms a basis of the set of functions, as is needed to decompose the Schrodinger picture fields $\phi(x)$ and $\pi(x)$.  For example, one may search for a completeness relation similar to (2.13) in Ref.~\cite{me2loop}.

With this motivation, let us consider the $\epsilon$ expansion of a new Ansatz
\beq
\phi(x,t)=\sum_{n=1}^\infty \epsilon^n f_n(\epsilon x,t) + \delta \sum_{n=1}^\infty \epsilon^n \g_{ n} (\epsilon x,t) \label{expl}
\eeq
which is suitable for long wavelength perturbations.  We expect $\g_B$ and $\g_T$ to be solutions.  Now the $O(\epsilon^n)$ terms are as above, but we will allow for different relativistic corrections to the frequency of the perturbations.  Therefore the $O(\delta\epsilon^n)$ terms will differ.

At order $O(\delta\epsilon)$ the equation of motion is just Eq.~(\ref{into1}).  However, we now write a solution, indexed by $k$, as
\beq
\g_1(\epsilon x,t)=G_k(\epsilon x)e^{-i(\Omega+\omega_k) t} +H_k(\epsilon x)e^{i(\Omega-\omega_k) t} \label{g1eq}
\eeq
for some $\omega_k$ which is equal to 
\beq
\omega_k=\sum_{n=1}^\infty \epsilon^n \omega_{k,n}.
\eeq
The solutions $\g_B$ and $\g_T$ would arise at $\omega_{k,n}=0.$  The $H_k$ term here plays the same role as in the homogeneous solution in $\g_2$ in the case of intermediate scale perturbations.  Here it is one order lower in $\epsilon$, appearing in $\g_1$ instead of $\g_2$, because two $x$ derivatives now lead to an additional power of $\epsilon$.

Using
\beq
(\Omega\pm \omega_k)^2=m^2\pm 2\epsilon m \omega_{k,1}+\epsilon^2(-1\pm 2m\omega_{k,2}+\omega^2_{k,1})
\eeq
one finds
\bea
\ddot \g_1(\epsilon x,t)&=&-\left[ m^2+ 2\epsilon m \omega_{k,1}+\epsilon^2(-1+ 2m\omega_{k,2}+\omega^2_{k,1}) \right] G_k(\epsilon x)e^{-i(\Omega+\omega_k) t}\\
&&-\left[ m^2- 2\epsilon m \omega_{k,1}+\epsilon^2(-1-2m\omega_{k,2}+\omega^2_{k,1}) \right] H_k(\epsilon x)e^{i(\Omega-\omega_k) t}.\nonumber
\eea

At order $O(\delta\epsilon^2)$ we arrive at
\bea
\ddot\g_2(\epsilon x,t)\oe{0}+m^2\g_2(\epsilon x,t)&=& 2 m \omega_{k,1}\left(
G_k(\epsilon x)e^{-i(\Omega+\omega_k) t}-H_k(\epsilon x)e^{i(\Omega-\omega_k) t}\right)\\
&&\hspace{-3cm}{ - } gV^{(3)}(0)\left[ F(\epsilon x)e^{i\Omega t}+F^*(\epsilon x)e^{-i\Omega t}
\right]\left[G_k(\epsilon x)e^{-i(\Omega+\omega_k) t} +H_k(\epsilon x)e^{i(\Omega-\omega_k) t}\right].\nonumber
\eea
Every term in the second line has a frequency of $O(\epsilon)$ or $\pm 2m+O(\epsilon)$ and so does not lead to a resonance.  On the other hand, the terms in the first line have frequency $\pm m+O(\epsilon)$ and so do  lead to a resonance unless $\omega_{k,1}=0$.  Therefore we fix $\omega_{k,1}=0$.

Again the homogeneous solution for $\g_2$ may be absorbed into $O(\epsilon)$ corrections to $G_k$ and $H_k$.  The particular solution is
\bea
\g_2(\epsilon x,t)&=&{} \frac{gV^{(3)}(0)}{3m^2}\left[-3\left(F(\epsilon x)G_k(\epsilon x)+F^*(\epsilon x)H_k(\epsilon x)\right){}e^{-i\omega_kt}\right.\\
&&\left.
+F^*(\epsilon x)G_k(\epsilon x)e^{-i(2\Omega+\omega_k)t}+F(\epsilon x)H_k(\epsilon x)e^{i(2\Omega-\omega_k)t}
\right].\nonumber
\eea

As in the usual $\epsilon$ expansion, the most interesting order has three powers of $\epsilon$, as that is the first place where spatial derivatives appear and so constraints may be placed on the spatial functions.  At $O(\delta\epsilon^3)$ one finds
\bea
\ddot \g_3(\epsilon x,t)\oe{0}+m^2 \g_3(\epsilon x,t)&=&
-\ddot\g_1(\epsilon x,t)|_{O(\epsilon^2)}-\ddot\g_2(\epsilon x,t)|_{O(\epsilon)}\label{pic3}\\
&&\hspace{-3cm}+\partial^2_{\epsilon { x }} \g_1(\epsilon x, t)-\frac{g^2V^{(4)}(0)}{2}f_1^2(\epsilon x,t)\g_1(\epsilon x,t)\nonumber\\
&&\hspace{-3cm}-{g V^{(3)}(0)} \left(f_1(\epsilon x,t)\g_2(\epsilon x,t)+f_2(\epsilon x,t)\g_1(\epsilon x,t)
\right)\nonumber
\\
&&\hspace{-3.5cm}\supset
(-1+2m\omega_{k,2})G_k(\epsilon x)e^{-i(\Omega+\ok{})t}+(-1-2m\omega_{k,2})H_k(\epsilon x)e^{i(\Omega-\ok{})t}\nonumber\\
&&\hspace{-3cm}+
G\pp_k(\epsilon x)e^{-i(\Omega+\omega_k) t} +H\pp_k(\epsilon x)e^{i(\Omega-\omega_k) t} \nonumber\\
&&\hspace{-3cm}
- \frac{g^2V^{(4)}(0)}{2}\left[
F(\epsilon x)e^{i\Omega t}+F^{ *}(\epsilon x)e^{-i\Omega t}
\right]^2\left[ G_k(\epsilon x)e^{-i(\Omega+\omega_k) t} +H_k(\epsilon x)e^{i(\Omega-\omega_k) t}
\right]\nonumber\\
&&\hspace{-3cm}{ - }\frac{g^2 V^{(3)2}(0)}{{ 3} m^2}\left[
F(\epsilon x)e^{i\Omega t}+F^{ *}(\epsilon x)e^{-i\Omega t}
\right]\nonumber\\
&&\hspace{-3cm}\ \ \times\left[-3\left(F(\epsilon x)G_k(\epsilon x)+F^*(\epsilon x)H_k(\epsilon x)\right){}e^{-i\omega_kt}\right.\nonumber\\
&&\hspace{-3cm}\left.\ \ \ \ \  +F^*(\epsilon x)G_k(\epsilon x)e^{-i(2\Omega+\omega_k)t}+F(\epsilon x)H_k(\epsilon x)e^{i(2\Omega-\omega_k)t}
\right]\nonumber\\
&&\hspace{-3cm} +\frac{g^2 V^{(3)2}(0)}{{ 6 }m^2}
\left[ G_k(\epsilon x)e^{-i(\Omega+\omega_k) t} +H_k(\epsilon x)e^{i(\Omega-\omega_k) t}
\right]\nonumber\\
&&\hspace{-3cm}\ \  \times\left[
-F^2(\epsilon x)e^{2i\Omega t}-F^{*2}(\epsilon x)e^{-2i\Omega t}+6|F(\epsilon x)|^2
\right]\nonumber
\eea
where the $\supset$ denotes that we dropped the $\ddot\g_2$ terms, as they do not contribute to any resonance.

The terms proportional to $e^{i(\Omega-\ok{})t}$ are
\beq
\ddot \g_3+m^2\g_3\supset e^{i(\Omega-\ok{})t}\left[-(1+2m\omega_{k,2})H_k(\epsilon x)+ H\pp_k(\epsilon x)+\lambda_F(F^2(\epsilon x)G_k(\epsilon x)+2|F(\epsilon x)|^2H_k(\epsilon x))
\right]
\eeq
while those proportional to $e^{-i(\Omega+\ok{})t}$ are
\beq
\ddot \g_3+m^2\g_3\supset e^{-i(\Omega+\ok{})t}\left[(-1+2m\omega_{k,2})G_k(\epsilon x)+ G\pp_k(\epsilon x)+\lambda_F(F^{*2}(\epsilon x)H_k(\epsilon x)+2|F(\epsilon x)|^2G_k(\epsilon x))
\right].
\eeq
We learn that the resonances vanish when two coupled, linear ordinary differential equations are satisfied
\bea
-(1+2m\omega_{k,2})H_k(\epsilon x)+ H\pp_k(\epsilon x)+\lambda_F (F^2(\epsilon x)G_k(\epsilon x)+2|F(\epsilon x)|^2H_k(\epsilon x))&=&0\label{coup}\\
(-1+2m\omega_{k,2})G_k(\epsilon x)+ G\pp_k(\epsilon x)+\lambda_F (F^{*2}(\epsilon x)H_k(\epsilon x)+2|F(\epsilon x)|^2G_k(\epsilon x))&=&0.\nonumber
\eea
Choosing $\theta=0$ or $\pi$, which is just an arbitrary choice of initial time
, this is
\bea
-(1+2m\omega_{k,2})H_k(\epsilon x)+ H\pp_k(\epsilon x)+2\sech^2 (\epsilon(x-x_0))(G_k(\epsilon x)+2H_k(\epsilon x))&=&0\label{esp}\\
(-1+2m\omega_{k,2})G_k(\epsilon x)+ G\pp_k(\epsilon x)+2\sech^2 (\epsilon(x-x_0))(2G_k(\epsilon x)+H_k(\epsilon x))&=&0.\nonumber
\eea

\subsection{Solutions of Eq.~(\ref{coup})}

Eq. (\ref{coup}) simplifies to a very elegant form if we define 
\beq
G_\pm(\epsilon x)=G_k(\epsilon x)\pm \frac{F^*(\epsilon x)}{F(\epsilon x)}H_k(\epsilon x) 
\eeq
Indeed, using 
\beq
\lambda_F|F(\epsilon x)|^2=2\sech^2 (\epsilon(x-x_0))
\eeq
one finds that the second equation plus (minus) $F^*/F$ times the first gives
\bea
2m\omega_{k,2}G_-(\epsilon x)-G_+(\epsilon x)+G\pp_+(\epsilon x)+6\sech^2 (\epsilon(x-x_0))G_+(\epsilon x)&=&0\label{pt2}\\
2m\omega_{k,2}G_+(\epsilon x)-G_-(\epsilon x)+G\pp_-(\epsilon x)+2\sech^2 (\epsilon(x-x_0))G_-(\epsilon x)&=&0.\nonumber
\eea
These are two coupled equations of the P\"oschl-Teller form, where the coupling is given by simply, constant coefficient terms.  

These functions may be decomposed in a basis of eigenfunctions of the corresponding ordinary P\"oschl-Teller problems
\beq
G_-=\ppin \hk c_\hk \g^{(1)}_\hk(x)\hsp
G_+=\ppin{\hk\p} d_{\hk\p} \g^{(2)}_{\hk\p}(x)
\eeq
where $\dint$ here represents a sum over all continuum and bound modes, including zero modes, of the corresponding problem.  More precisely, where the index  $\hk$ may be a continuum mode or $B$ while $\hk\p$ may be a continuum mode, $B$ or $S$. 

The modes satisfy the Sturm-Liouville problems
\bea
&&\left(\partial^2_{\epsilon x}+2\sech^2 (\epsilon x)\right)\g^{(1)}_\hk(x)=-\frac{\hk^2}{\epsilon^2} \g^{(1)}_\hk(x)\hsp 
\left(\partial^2_{\epsilon x}+6\sech^2 (\epsilon x)\right)\g^{(2)}_{\hk\p}(x)=-\frac{\hk^{\prime 2}}{\epsilon^2} \g^{(2)}_{\hk\p}(x)\nonumber\\
&&\left(\partial^2_{\epsilon x}+6\sech^2 (\epsilon x)\right)\g^{(2)}_B(x)=4 \g^{(2)}_B(x)\hsp
\left(\partial^2_{\epsilon x}+6\sech^2 (\epsilon x)\right)\g^{(2)}_S(x)= \g^{(2)}_S(x)\nonumber\\
&&\left(\partial^2_{\epsilon x}+2\sech^2 (\epsilon x)\right)\g^{(1)}_B(x)= \g^{(1)}_B(x)
\eea
where for concreteness we have set $x_0=0$.

Inserting this basis into the equations of motion Eq.~(\ref{pt2}) one finds
\bea
2m\omega_{k,2}\ppin \hk c_\hk \g^{(1)}_\hk(x)&=&-3d_B\g^{(2)}_B(x)+\pin{\hk\p}d_{\hk\p} \frac{\nu^2_{\hk\p}}{\epsilon^2} \g^{(2)}_{\hk\p}(x)\label{aeom}\\
2m\omega_{k,2}\ppin {\hk\p} d_{\hk\p} \g^{(2)}_{\hk\p}(x)&=&\pin{\hk}c_\hk  \frac{\nu^2_{\hk}}{\epsilon^2}\g^{(1)}_\hk(x)\nonumber
\eea
where we have defined
\beq
\nu_{k}=\sqrt{\epsilon^2+k^2}.
\eeq
Note that $\nu$ is not the frequency of any particle, as $\epsilon$ is not a mass, rather it is a kind of effective energy for a long wavelength perturbation of the oscillon.   The sums $\dint$ on the left hand side contain all bound modes.  However if $\omega_{k,2}=0$ then the left hand side vanishes and so $d_S$ and $c_B$, the coefficients of $\g_S^{(2)}$ and $\g_B^{(1)}$, do not appear in the equations.  As a result, $G_-=\g_B^{(1)}$ and $G_+=\g_S^{(2)}$ are zero-mode solutions.

\subsubsection{Zero Modes}

Consider the special case $\omega_{k,2}=0$.  In this case, these coupled equations are easily solved. Indeed, the equations decouple 
\beq
-G_+(\epsilon x)+G\pp_+(\epsilon x)+6\sech^2 (\epsilon(x-x_0))G_+(\epsilon x)=0
\eeq
\beq
-G_-(\epsilon x)+G\pp_-(\epsilon x)+2\sech^2 (\epsilon(x-x_0))G_-(\epsilon x)=0.
\eeq
These are the $\lambda=2$  and $\lambda=1$ P\"oschl-Teller equations with energy $-1/2$.  This energy corresponds to the highest bound state in each case.

In other words, if $G_-=0$ then $G_+$ is the higher of the two bound states, and, using the known solution, one finds that  $\g=\g_T$ is the temporal zero mode.  On the other hand, if $G_+=0$ then $G_-$ is the only bound state, and it yields the spatial zero mode $\g=\g_B$.

\subsection{Near Threshold Continuum Modes}

\subsubsection{Continuum Modes}

Consider $\epsilon x\gg 1/m$, so that the $F^2$ terms in Eq.~(\ref{coup}) vanish
\beq
-(1+2m\omega_{k,2})H_k(\epsilon x)+ H\pp_k(\epsilon x)=0\hsp
(-1+2m\omega_{k,2})G_k(\epsilon x)+ G\pp_k(\epsilon x)=0.
\eeq
We expect, based on intuition with time-independent backgrounds, that only negative frequency modes $\ok{}>0$ will be interesting.  Therefore we restrict our attention to $\omega_{k,2}>0$.  Then we see that $H_k$ always falls off exponentially far from the oscillon, however $G_k$ becomes a plane wave for $\omega_{k,2}>1/(2m)$, corresponding to $\Omega+\ok{}>m$ and so the frequency of $\g_1$ beyond the mass threshold.  It remains to check that these lead to consistent solutions to (\ref{coup}).

\begin{figure}[htbp]
\centering
\includegraphics[width = 0.45\textwidth]{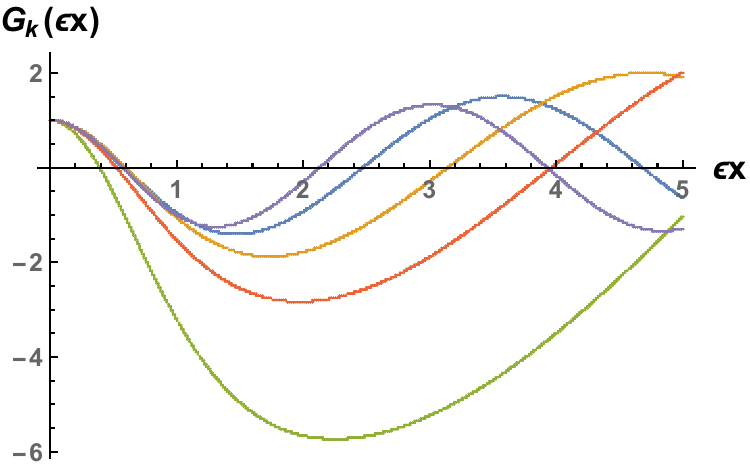}
\includegraphics[width = 0.45\textwidth]{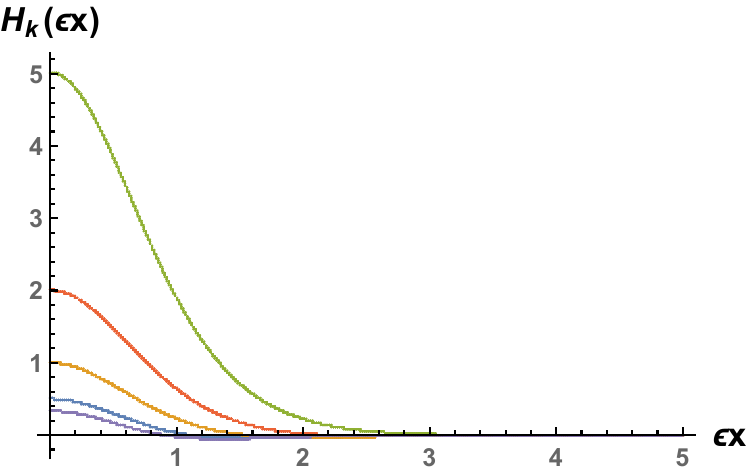}
\caption{Numerical solutions for even continuum perturbations.  The functions $G_k(\epsilon x)$ and $H_k(\epsilon x)$, which represent the coefficients of the positive and negative frequency parts of the perturbations respectively, are plotted at $2m\omega_{k,2}$ equal to 1.2 (green), 1.5 (red), 2 (orange), 3 (blue) and 4 (purple).
}\label{ghfig}
\end{figure}

\begin{figure}[htbp]
\centering
\includegraphics[width = 0.45\textwidth]{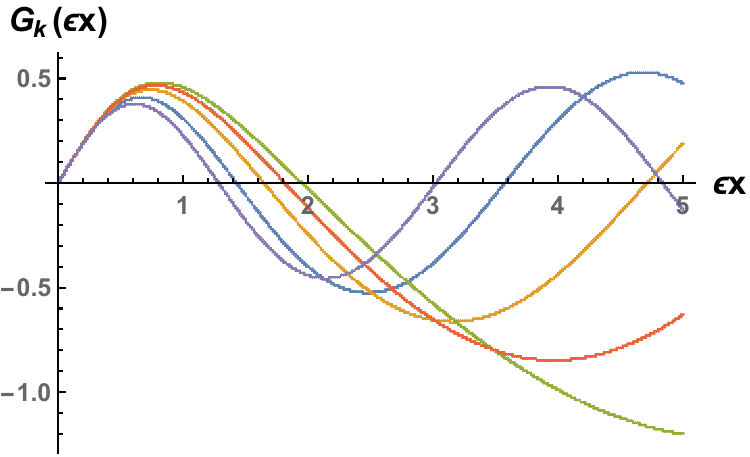}
\includegraphics[width = 0.45\textwidth]{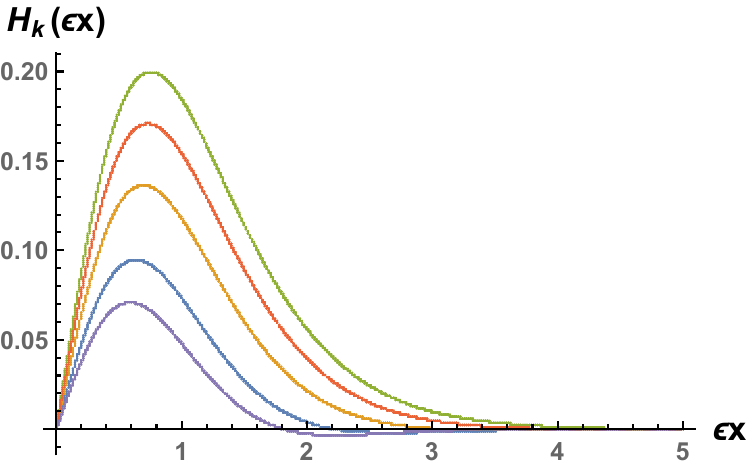}
\caption{Numerical solutions for odd continuum perturbations, with conventions as in Fig.~\ref{ghfig}.
}\label{ghofig}
\end{figure}

When $\omega_{k,2}>1/(2m)$, $G$ is continuous and $H$ is bound, so in principle for a given $\omega_{k,2}$ there may be several solutions corresponding to different numbers of nodes of $H$.   We have numerically obtained a branch of solutions for all frequencies $\omega_{k,2}$ in this regime, some of which are plotted in Figs.~\ref{ghfig} and \ref{ghofig} at $x_0=0$.

At higher frequencies, $H_k$ becomes smaller.  This is consistent with the fact that at very lower frequencies, in Sec.~\ref{intsez}, we saw in Eq.~(\ref{sop}) that it is suppressed.  Note that in the case of the even perturbations the $y$-intercepts of $H/G$ are, within numerical errors, equal to
\beq
\frac{H_k(0)}{G_k(0)}=\frac{1}{2m\omega_{k,2}-1}.
\eeq
Indeed, close to the threshold at $\omega_{k,2}=1/(2m)$, we find that $H_k(\epsilon x)/G_k(0)$ tends to $\sech^2(\epsilon x)/(2m\omega_{k,2}-1)$.  Similarly, in the case of the odd functions, the slopes at the origin are related by
\beq
\frac{H\p_k(0)}{G\p_k(0)}=\frac{1}{2m\omega_{k,2}+1}.
\eeq
This gives hope that analytic solutions may be found.  We also note that the even and odd solutions appear to be 90 degrees out of phase far from the oscillon, and so may be assembled into $e^{ikx}$ or $e^{-ikx}$, suggesting that reflectionless oscillon-radiation scattering persists even at these long wavelengths.  

\section{Nonperiodic Solutions} \label{movsez}

\subsection{A Moving Oscillon}
We have not yet found all solutions $\g(x,t)$ of Eq.~(\ref{sl}), as we have assumed that $\g(x,t+2\pi/\Omega)$ equals $\g(x,t)$ up to a phase.  Another solution is provided by a moving oscillon.  In the defining frame, this can be constructed by simply boosting the oscillon solution
\beq
\phi(x,t)=f(x-vt,t-vx).
\eeq
In the comoving frame this is
\beq
\phi(x,t)=v\g_M(x,t)=f(x-vt,t-vx)-f(x,t)=v\left(t\g_B(x,t)+x\g_T(x,t)\right).
\eeq

Recalling that $\g_B(x,t)$ and $\g_T(x,t)$ are solutions of (\ref{sl}), we can check that $\g_M(x,t)$ is a solution
\bea
\left(\partial_x^2-V\pp(g f(\epsilon x,t))-\partial^2_t\right)\g_M(x,t)&=&\left[\left(\partial_x^2-V\pp(g f(\epsilon x,t))-\partial^2_t\right),t\right]\g_B(x,t)\\
&&\hspace{-6.5cm}+t\left(\partial_x^2-V\pp(g f(\epsilon x,t))-\partial^2_t\right)\g_B(x,t)+x\left(\partial_x^2-V\pp(g f(\epsilon x,t))-\partial^2_t\right)\g_T(x,t)\nonumber\\
&&\hspace{-6.5cm}+\left[\left(\partial_x^2-V\pp(g f(\epsilon x,t))-\partial^2_t\right),x\right]\g_T(x,t)=-2\dot\g_B(x,t)+2\g\p_T(x,t)=(2-2)\dot f\p(x,t)=0.\nonumber
\eea

Note that this solution corresponds to the initial conditions
\beq
\g_M(x,0)=x\g_T(x,0)
\hsp\dot\g_M(x,0)=\g_B(x,0)+x\dot\g_T(x,0).
\eeq
In the case of the moving kink, $\g_T=-\dot f=0$ and so only the $\g_B$ term in $\dot\g_M$ appears.  Returning to the general oscillon case, instead of periodicity, $\g_M(x,t)$ satisfies the condition
\bea
\g_M\left(x,t+\frac{2\pi}{\Omega}\right)&=&\left(t+\frac{2\pi}{\Omega}\right)\g_B\left(x,t+\frac{2\pi}{\Omega}\right)+x\g_T\left(x,t+\frac{2\pi}{\Omega}\right)\label{moveper}\\
&=&\left(t+\frac{2\pi}{\Omega}\right)\g_B\left(x,t\right)+x\g_T\left(x,t\right)=\g_M\left(x,t\right)+\frac{2\pi}{\Omega}\g_B(x,t).\nonumber
\eea

\subsection{The Monodromy Matrix}
Following the general treatment of perturbations $\g$ of stable orbits $f(x,t)$ in Ref.~\cite{pars}, we may define a monodromy matrix $M$ such that
\beq
M\left(
\begin{tabular}{c}
$\g(x,0)$\\
$\dot\g(x,0)$
\end{tabular}
\right)
=\left(\begin{tabular}{c}
$\g(x,2\pi/\Omega)$\\
$\dot\g(x,2\pi/\Omega)$
\end{tabular}
\right).
\eeq
The vectors $(\g_k(x,0),\dot\g_k(x,0))$ are eigenvectors, as are $(\g^*_k(x,0),\dot\g^*_k(x,0))$
\beq
M\left(
\begin{tabular}{c}
$\g_k(x,0)$\\
$\dot\g_k(x,0)$
\end{tabular}
\right)
=e^{-2\pi i\ok{}/\Omega}\left(
\begin{tabular}{c}
$\g_k(x,0)$\\
$\dot\g_k(x,0)$
\end{tabular}
\right)\hsp
M\left(
\begin{tabular}{c}
$\g^*_k(x,0)$\\
$\dot\g^*_k(x,0)$
\end{tabular}
\right)
=e^{2\pi i\ok{}/\Omega}\left(
\begin{tabular}{c}
$\g^*_k(x,0)$\\
$\dot\g^*_k(x,0)$
\end{tabular}
\right)
\eeq
where the second equation is the complex conjugate of the first, using the fact that $M$ is real.  Similarly the zero-modes are preserved by $M$
\beq
M\left(
\begin{tabular}{c}
$\g_B(x,0)$\\
$\dot\g_B(x,0)$
\end{tabular}
\right)
=\left(
\begin{tabular}{c}
$\g_B(x,0)$\\
$\dot\g_B(x,0)$
\end{tabular}
\right)\hsp
M\left(
\begin{tabular}{c}
$\g_T(x,0)$\\
$\dot\g_T(x,0)$
\end{tabular}
\right)
=\left(
\begin{tabular}{c}
$\g_T(x,0)$\\
$\dot\g_T(x,0)$
\end{tabular}
\right).
\eeq
Finally Eq.~(\ref{moveper}) implies
\beq
M\left(
\begin{tabular}{c}
$\g_M(x,0)$\\
$\dot\g_M(x,0)$
\end{tabular}
\right)
=\left(
\begin{tabular}{c}
$\g_M(x,0)+(2\pi/\Omega)\g_B(x,0)$\\
$\dot\g_M(x,0)+(2\pi/\Omega)\dot\g_B(x,0)$
\end{tabular}
\right).
\eeq

Note that $M$ is not diagonalizable, in fact on the basis $(\g_B,\g_M)$ it is upper-triangular.  Nonetheless, the monodromy matrix acts on all perturbations and so vectors $(\g(x,0),\dot\g(x,0))$ span the space of pairs of functions, corresponding to arbitrary initial conditions.  The pairs $(\g(x,0),\dot\g(x,0))$ corresponding to $\g_k,\ \g_B,\ \g_T$ and $\g_M$ are linearly independent and, if they span the space of solutions of (\ref{sl}), then these pairs are a basis of the space of pairs of functions.  This means that they can be used to decompose Schrodinger picture pairs $(\phi(x),\pi(x))$ in quantum field theory and so will allow a quantization of the oscillon following the strategy of Refs.~\cite{cahill76,mekink}.

\section{Remarks}

In classical field theory, oscillons dominate configurations fairly ubiquitously \cite{gleiser94,gani21,li24, J, Simas} after violent events have excited all perturbations, and short-lived perturbations have had time to dissipate.  As a result, oscillons play a key role in many phenomenological contexts \cite{lozanov23,muia23,pirvu23,corral24}.  It is often claimed that this is not the case in the real world, as the quantum oscillons of Refs.~\cite{hertzberg,tanmay,tanmay21} emit radiation with such a high luminosity, of order the meson mass squared, that, if they continued to radiate, they would dissipate rather quickly.

The key question becomes, what is the end point of this decay?  Is it pure radiation or a more stable quantum oscillon?  This motivates us to search for a more stable quantum oscillon than that of Refs.~\cite{hertzberg,tanmay}, see e.e., ~\cite{Q-osc} and thus our interest in the quantization of oscillon.

We wish not only to find the quantum corrections to the oscillon mass, as in the treatment of breathers in Ref.~\cite{dhnsg}, but we also desire to explicitly find the oscillon state.  This will allow us to answer questions such as whether the phase $\theta$ can be fixed for a quantum oscillon, or rather whether it is necessarily described by some continuous wave functions.  Or, more likely, whether it smears over a time scale which is suppressed in powers of the coupling $g$.  To do this, we will use the approach of Refs.~\cite{cahill76,mekink,me2loop}.  But this approach, as well as that of Ref.~\cite{dhnsg}, requires the full monodromy matrix.  In particular, one needs to find explicit expressions for a set of perturbations which spans the space of initial conditions, consisting of all possible perturbations of the field and its first time derivative.  In the case of the quantum kink, the perturbations were described by a Sturm-Liouville equation, whose solutions are known to provide a basis.  However in the present case, we have not yet shown completeness for our solutions.  Perhaps this needs to be done numerically.  Nonetheless, numerically we have searched and we have not found other solutions.

In summary, we have classified the perturbations of the small amplitude oscillon and we have numerical evidence that our classification is complete.  We stress that our analysis applies to models with any potential which admits a small amplitude oscillon solution, which in particular requires that the mass threshold be nonzero.  If the solutions found here are indeed complete, then we are now ready to quantize the oscillon.  

Interestingly, the non-existence of massive bound modes for oscillons provides a further support for the recently proposed explanation of amplitude modulation. A genuine oscillon enjoys two independent degrees of freedom: the fundamental frequency $\Omega$ and a frequency of the amplitude modulations. In \cite{BRSW} it has been argued that modulated oscillon can be viewed as a semi-bound state of two (or more) unmodulated, fundamental oscillons. Potentially, one might imagine another explanation where the one parameter $\epsilon$ expansion \cite{fodor} gives the fundamental frequency while a linear perturbation leads to an internal vibrational mode which is the second degree of freedom needed for the modulations. Since small amplitude oscillons do not host vibrational modes this possibility is excluded. 

\section*{Acknowledgements}

KS acknowledge financial support from the Polish National Science Centre 
(Grant NCN 2021/43/D/ST2/01122).


\begin{thebibliography}{99}

\bibitem{osc76}
I.~L.~Bogolyubsky and V.~G.~Makhankov,
``Lifetime of Pulsating Solitons in Some Classical Models,''
Pisma Zh. Eksp. Teor. Fiz. \textbf{24} (1976), 15-18

\bibitem{gleiser94}
M.~Gleiser,
``Pseudostable bubbles,''
Phys. Rev. D \textbf{49} (1994), 2978-2981
doi:10.1103/PhysRevD.49.2978
[arXiv:hep-ph/9308279 [hep-ph]].

\bibitem{G-cosm} M. Gleiser, Oscillons in scalar field theories: Applications in higher dimensions and inflation, Int. J. Mod. Phys. D {\bf 16} (2007) 219.

\bibitem{GS} M. Gleiser and D. Sicilia, A General Theory of Oscillon Dynamics, Phys. Rev. D 80 (2009) 125037.


\bibitem{ks74}
A. M.  Kosevich and A. S. Kovalev,
``Self-localization of vibrations in a one-dimensional anharmonic
chain."
Zh. Eksp. Teor. Fiz. \textbf{67} (1974) 1793-1804


\bibitem{fodor}
G.~Fodor, P.~Forgacs, Z.~Horvath and A.~Lukacs,
``Small amplitude quasi-breathers and oscillons,''
Phys. Rev. D \textbf{78} (2008), 025003
doi:10.1103/PhysRevD.78.025003
[arXiv:0802.3525 [hep-th]].

\bibitem{pars}
L. A. Pars,
``A Treatise on Analytical Dynamics,''
(Heinemann, London, 1965)

\bibitem{dhnsg}
R.~F.~Dashen, B.~Hasslacher and A.~Neveu,
``The Particle Spectrum in Model Field Theories from Semiclassical Functional Integral Techniques,''
Phys. Rev. D \textbf{11} (1975), 3424
doi:10.1103/PhysRevD.11.3424

\bibitem{dhn2}
  R.~F.~Dashen, B.~Hasslacher and A.~Neveu,
  ``Nonperturbative Methods and Extended Hadron Models in Field Theory 2. Two-Dimensional Models and Extended Hadrons,''
  Phys.\ Rev.\ D {\bf 10} (1974) 4130.
 doi:10.1103/PhysRevD.10.4130

\bibitem{wang22}
Y.~J.~Wang, Q.~X.~Xie and S.~Y.~Zhou,
``Excited oscillons: cascading levels and higher multipoles,'' Phys. Rev. D {\bf 108} (2023) 025006.

\bibitem{van23}
F.~van Dissel, O.~Pujolas and E.~Sfakianakis,
``Oscillon spectroscopy,'' JHEP {\bf 07} (2023) 194.

\bibitem{amin10}
M.~A.~Amin,
``Inflaton fragmentation: Emergence of pseudo-stable inflaton lumps (oscillons) after inflation,''
[arXiv:1006.3075 [astro-ph.CO]].

\bibitem{tesi}
F. Van Dissel,
``Multi-Component Oscillons,"
Masters Thesis at Leiden University, 2020,
https://hdl.handle.net/1887/135806.

\bibitem{BRSW} F. Blaschke, T. Romanczukiewicz, K. Slawinska, and A. Wereszczynski, "Amplitude modulations and resonant decay of excited oscillons",  Phys. Rev. E {\bf 110} (2024) 014203 [arXiv:2403.00443 [hep-th]]. 

\bibitem{gleiser2024}
M. Gleiser and  D. Sicilia, 
''Analytical Characterization of Oscillon Energy and Lifetime''
Phys. Rev. Lett. \textbf{101}, 011602
doi.org/10.1103/PhysRevLett.101.011602
[arXiv:0804.0791 [hep-th]]

\bibitem{copeland1995}
E. J. Copeland, M. Gleiser, and H.-R. Müller
''Oscillons: Resonant configurations during bubble collapse''
Phys. Rev. D \textbf{52}, 1920
doi.org/10.1103/PhysRevD.52.1920
[arXiv:hep-ph/9503217 [ht-ph]]

\bibitem{For} G. Fodor, P. Forgacs, Z. Horvath, and Mark Mezei, "Computation of the radiation amplitude of oscillons", Phys. Rev. D {\bf 79} (2009) 065002.

\bibitem{DRSW} P. Dorey, T. Romanczukiewicz, Y. Shnir, A. Wereszczynski, "Oscillons in gapless theories", Phys. Rev. D {\bf 109} (2024) 085017 [arXiv:2312.05308 [hep-th]]. 

\bibitem{flugge}
S. Fl\"ugge,
``Practical Quantum Mechanics,"
Springer-Verlag Berlin Heidelberg (1999),
doi:10.1007/978-3-642-61995-3


\bibitem{me2loop}
J.~Evslin and H.~Guo,
``Two-Loop Scalar Kinks,''
Phys. Rev. D \textbf{103} (2021) no.12, 125011
doi:10.1103/PhysRevD.103.125011
[arXiv:2012.04912 [hep-th]].




\bibitem{cahill76}
K.~E.~Cahill, A.~Comtet and R.~J.~Glauber,
``Mass Formulas for Static Solitons,''
Phys. Lett. B \textbf{64} (1976), 283-285
doi:10.1016/0370-2693(76)90202-1


\bibitem{mekink}
J.~Evslin,
``Manifestly Finite Derivation of the Quantum Kink Mass,''
JHEP \textbf{11} (2019), 161
doi:10.1007/JHEP11(2019)161
[arXiv:1908.06710 [hep-th]].

\bibitem{gani21}
V.~A.~Gani, A.~M.~Marjaneh and K.~Javidan,
``Exotic final states in the $\varphi ^8$ multi-kink collisions,''
Eur. Phys. J. C {\bf 81} (2021) no.12, 1124
doi:10.1140/epjc/s10052-021-09935-7
[arXiv:2106.06399 [hep-th]].



\bibitem{li24}
X.~Li and L.~Long,
``Radiation-like Shock Waves in Kink Scattering,''
[arXiv:2407.14479 [hep-th]].

\bibitem{J} S. Navarro-Obregon, L. M. Nieto, and J. M. Queiruga, Inclusion of radiation in the CCM approach of the $\phi^4$ model, Phys. Rev. E {\bf 108} (2023) 044216. 
\bibitem{Simas} F. C. Lima, F. C. Simas, K. Z. Nobrega, and A. R. Gomes,
Scattering of metastable lumps in a model with a false vacuum,
Phys. Lett. B {\bf 822} (2021) 136707.

\bibitem{lozanov23}
K.~D.~Lozanov, M.~Sasaki and V.~Takhistov,
``Universal Gravitational Wave Signatures of Cosmological Solitons,''
[arXiv:2304.06709 [astro-ph.CO]].

\bibitem{muia23}
J.~C.~Aurrekoetxea, K.~Clough and F.~Muia,
``Oscillon formation during inflationary preheating with general relativity,''
Phys. Rev. D {\bf 108} (2023) 023501 .

\bibitem{pirvu23}
D.~P\^\i{}rvu, M.~C.~Johnson and S.~Sibiryakov,
``Bubble velocities and oscillon precursors in first order phase transitions,''
[arXiv:2312.13364 [hep-th]].

\bibitem{corral24}
D.~del-Corral,
``Self-resonance after inflation: The case of $\alpha$-attractor models,''
[arXiv:2406.04017 [hep-th]].

\bibitem{hertzberg}
M.~P.~Hertzberg,
``Quantum Radiation of Oscillons,''
Phys. Rev. D \textbf{82} (2010), 045022
doi:10.1103/PhysRevD.82.045022
[arXiv:1003.3459 [hep-th]].

\bibitem{tanmay}
J.~Oll\'e, O.~Pujolas, T.~Vachaspati and G.~Zahariade,
``Quantum Evaporation of Classical Breathers,''
Phys. Rev. D \textbf{100} (2019) no.4, 045011
doi:10.1103/PhysRevD.100.045011
[arXiv:1904.12962 [hep-th]]

\bibitem{tanmay21}
M.~Mukhopadhyay, E.~I.~Sfakianakis, T.~Vachaspati and G.~Zahariade,
``Kink-antikink scattering in a quantum vacuum,''
JHEP \textbf{04} (2022), 118
doi:10.1007/JHEP04(2022)118
[arXiv:2110.08277 [hep-th]].

\bibitem{Q-osc} J. Evslin, T. Romanczukiewicz and A. Wereszczynski, "Quantum Oscillons May be Long-Lived", JHEP {\bf 08} (2023) 182 [arXiv:2305.18056 [hep-th]]. 

\end{thebibliography}
\end{document}